\documentclass[11pt]{article}
\RequirePackage[OT1]{fontenc}
\usepackage{amsthm,amsmath,amssymb,amsfonts,bm,enumerate,xcolor,graphicx,natbib,color,url}
\usepackage{bbm}
\usepackage{dsfont}
\RequirePackage[colorlinks,citecolor=blue,urlcolor=blue]{hyperref}
\usepackage{hyperref}
\usepackage{subcaption} 
\usepackage{pifont}
\usepackage{textcmds}
\usepackage{booktabs}
\usepackage{enumitem}
\usepackage{multirow}
\usepackage{booktabs}
\usepackage{adjustbox}
\usepackage[usestackEOL]{stackengine}
\usepackage[toc,page]{appendix} 
\usepackage{algorithm,algorithmic}
\usepackage{lipsum}
\usepackage{setspace}

\usepackage{setspace}
\usepackage{multirow}

\usepackage{pgfplots}
\usepackage{tikz}
\usetikzlibrary{shapes, arrows}

\tikzstyle{connector} = [draw, -latex']

\usepackage{ulem} 
\def\AH#1{{\textcolor{red}{#1}}}

\def\AH#1{\footnote{{\textcolor{red}{Arnab\AH{}: #1}}}}
\def\AH#1{{\textcolor{red}{(Arnab: #1)}}}

\usepackage[english]{babel}
\usepackage{amsfonts}
\usepackage{amssymb}
\usepackage{graphicx}
\usepackage{enumerate}

\newcommand{\blind}{1}

\pgfplotsset{compat=1.18}
\begin{document}
\thispagestyle{empty}
\baselineskip=28pt
\vskip 2mm


 \begin{center} {\Large{\bf Deep graphical regression for jointly  moderate and extreme Australian wildfires}}
\end{center}

\baselineskip=12pt

\vskip 1mm

\if1\blind
{
\begin{center}
\large
Daniela Cisneros$^1$, Jordan Richards$^1$, Ashok Dahal$^2$, \\Luigi Lombardo$^2$, and Rapha\"el Huser$^1$\\ 
\end{center}
\footnotetext[1]{
\baselineskip=10pt Statistics Program, Computer, Electrical and Mathematical Sciences and Engineering (CEMSE) Division, King Abdullah University of Science and Technology (KAUST), Thuwal 23955-6900, Saudi Arabia. 

E-mail: jordan.richards@kaust.edu.sa}
\footnotetext[2]{
\baselineskip=10pt University of Twente, Faculty of Geo-Information Science and Earth Observation (ITC), PO Box 217, Enschede, AE 7500, Netherlands.}
} \fi

\baselineskip=26pt
\vskip 2mm
\centerline{\today}
\vskip 2mm


{\large{\bf Abstract}} 
Recent wildfires in Australia have led to considerable economic loss and property destruction, and there is increasing concern that climate change may exacerbate their intensity, duration, and frequency. Hazard quantification for extreme wildfires is an important component of wildfire management, as it facilitates efficient resource distribution, adverse effect mitigation, and recovery efforts. However, although extreme wildfires are typically the most impactful, both small and moderate fires can still be devastating to local communities and ecosystems. Therefore, it is imperative to develop robust statistical methods to reliably model the full distribution of wildfire spread. We do so for a novel dataset of Australian wildfires from 1999 to 2019, and analyse monthly spread over areas approximately corresponding to Statistical Areas Level~1 and~2 (SA1/SA2) regions. Given the complex nature of wildfire ignition and spread, we exploit recent advances in statistical deep learning and extreme value theory to construct a parametric regression model using graph convolutional neural networks and the extended generalized Pareto distribution, which allows us to model wildfire spread observed on an irregular spatial domain. We highlight the efficacy of our newly proposed model and perform a wildfire hazard assessment for Australia and population-dense communities, namely Tasmania, Sydney, Melbourne, and Perth.



\baselineskip=26pt



\baselineskip=26pt

\section{Introduction}

Due to their unprecedented scale and severity, the infamous Black Summer bushfires, which occurred from late 2019 to early 2020, were a significant environmental disaster. Their effects are believed to have been heavily exacerbated by climate change \citep{ward2020impact,haque2021wildfire}. Destruction of property on a massive scale, with an estimated 2000 homes destroyed and billions of dollars in damages, motivates a real need for informed urban planning and updates to both building codes and disaster management strategies, with a view towards minimizing losses and maximizing resilience in wildfire-prone areas \citep{gibbons2012land,price2012efficacy}. Such extreme wildfires have negative impacts on human health via the facilitation of hazardous air quality levels, which can lead to respiratory issues in vulnerable populations \citep{finlay2012health}. Aside from their human impact, the Black Summer fires had devastating impacts on local wildlife and biodiversity, with an estimated excess of 3 billion animals thought to have been affected through habitat loss \citep{levin2021unveiling}. In order to mitigate the impacts of future, similarly devastating events, there is a need to develop statistical models that can perform robust wildfire hazard assessments in Australia. To that end, we build a hybrid statistical deep-learning model for the occurrence and spread of Australian wildfires. We utilize this model to build continental hazard maps that facilitate hazard evaluation of population-dense communities.

 Studies of wildfires often make use of either point-pattern \citep{genton2006spatio, hering2009modeling, juan2012pinpointing} or areal \citep{serra2012spatio, roger2015rinla} datasets. Whilst the former extract information from the (random) location of wildfire occurrences, it can be difficult to build parsimonious models for them without relying on complex point process models \citep[see, e.g.,][]{koh2021spatiotemporal}. By contrast, areal datasets, that describe wildfire characteristics aggregated over space-time polygons, may be better suited to facilitate hazards assessment as they describe the characteristics of wildfires in a fixed frame of reference. Whilst many areal datasets of Australian burnt areas exist \citep{matthews2012field, GIGLIO201872, BOUGUETTAYA2022108309}, they are often criticised as the burnt area is aggregated over regular spatial gridboxes. If one were to perform a hazard analysis with these data, it would not truly reflect the hazard to local populations and communities, as these are not arranged in a regular fashion. In this paper, we construct a novel dataset of aggregated burnt areas over irregular spatial regions fabricated from the Level~1 and 2 Australian Statistical Geography Standard (ASGS) boundaries \citep{australian2011australian}. The smaller Level~2 boundaries are utilised in the population-dense ``greater'' capital city areas, that is, Sydney, Melbourne, Brisbane, Perth, Darwin, Hobart, and Adelaide, with the coarser Level~1 boundaries adopted elsewhere.
 
Statistical approaches for modelling wildfires have been proposed, with a particular reliance on univariate probability models, see, for example, \cite{cruz2012anatomy}, \cite{nhess-15-417-2015}, \cite{rios2018studying}, \cite{joseph2019spatiotemporal}, \cite{abram2021connections}, and \cite{xi2019statistical}. Several studies have focused on modeling Australian wildfires, particularly their extremes, using regression frameworks and have exploited generalized linear models \citep{kumar2008mapping, collins2022warmer}, generalized additive models \citep{storey2022prediction}, machine-learning \citep{wang2022extreme}, and extreme value theory \citep{douglas2014use, cisneros2023spatial}, with the latter adopting the generalized Pareto distribution (GPD) to model extreme wildfire severity. This distribution has also been successfully exploited for modelling wildfires across other spatial domains; see, for example, \cite{de2009spatial}, \cite{mendes2010spatial}, \cite{li2021temporal}, \cite{koh2021spatiotemporal}, and \cite{richards2022insights}. 

GPDs are asymptotically-justified by extreme-value theory and are thus often used to estimate quantiles far into the upper tails of a distribution \citep{coles2001introduction}. They are typically exploited in a peaks-over-threshold framework, and fitted only to excesses above a pre-specified threshold, with empirical estimates of the distribution used to model the distribution below the threshold; see \cite{coles2001introduction} for details. As such, they do not provide a characterisation of the behaviour of non-extreme (i.e., low or moderate) events. Moreover, their inference may be sensitive to the choice of threshold, which can be problematic in practice \citep{Naveau.etal:2016}.  Modeling non-extreme, as well as extreme, wildfires is still crucial for hazard analysis, as they can have significant impacts on local ecosystems, homes, and communities, and can quickly develop into extreme and devastating wildfires. This is particularly relevant for Australia, where wildfires are a common occurrence \citep{valente2021spatio}. Whilst the GPD can be fitted to the entire range of data, a more appropriate model may be one that has asymptotically-justified GPD upper-tails but a different parametric characterisation for the bulk of the distribution. Mixture distributions that can capture both the upper-tails and bulk of data have been constructed by stitching together GPDs with parametric distributions, see, for example, \cite{Frigessi.etal:2003} and \cite{carreau2009hybrid}, but these models often place restrictive constraints on the parameter space that make their inference difficult. More recently, there have been extensive research efforts to develop models with more flexible lower-tail, bulk, and upper-tail behaviours \citep[e.g.,][]{yadav2021spatial,Stein2021,stein2021parametric}. In particular,  \cite{Papastathopoulos.Tawn:2013} and \cite{Naveau.etal:2016} propose the extended generalized Pareto distribution (eGPD) as a flexible alternative to mixture distributions, which addresses both their limitations and those of standard GPD models. The eGPD has been successfully applied in the context of precipitation \citep{Naveau.etal:2016,tencaliec2020flexible, de2021extreme, haruna2022modeling} and landslide \citep{yadav2022joint} modelling, but its potential for modelling wildfires has not yet been fully exploited. We adopt the eGPD as our distributional model for wildfire spread.

Machine learning methods have shown promising success in the modelling of wildfire spread and susceptibility, see, for example, \cite{radke2019firecast}, \cite{zhang2019graph}, \cite{bergado2021predicting}, and \cite{cisneros2021combined,cisneros2023spatial}, as well as the propagation of wildfire fronts \citep{dabrowski2023bayesian,yoo2023using}. Hence, we leverage their predictive power to model wildfire spread through a parametric regression framework. In particular, we adopt a deep regression framework, as neural networks are capable of capturing the highly complex structure that we expect to be exhibited by the processes that drive Australian wildfire ignition and spread \citep{tolhurst2008phoenix}. Deep extreme value regression models have been successfully implemented by, for example,  \cite{carreau2009hybrid}, \cite{carreau2011stochastic}, \cite{rietsch2013network}, and \cite{pasche2022neural}, but they adopt neural network architectures that do not exploit spatial structure in data.  \cite{richards2022unifying} and \cite{richards2022insights} advocate the use of convolution neural networks (CNNs), see review by \cite{Gallego2022}, to capture complex spatial structure in the predictor variables when they model U.S.\ and European wildfires, respectively. Both studies illustrate that better fitting models for extreme wildfire spread, relative to those that use standard densely-connected neural networks, can be produced by exploiting spatial structure in the covariates. A crucial drawback of CNNs is that they are only applicable to regularly gridded data, which is a property not exhibited by the data that we use in our study. We instead develop a graphical regression model using graph convolutional neural networks (GCNNs), see \cite{kipf2016semi}, which extend the notion of convolutions in CNNs from regularly gridded to irregularly gridded or graphical input data.

The rest of the paper is organized as follows. In Section~\ref{sec:DataDescr}, we introduce the novel dataset of burnt areas resulting from Australian wildfires that we use in our wildfire hazard assessment study and we discuss its construction. In Section~\ref{sec:Methodology}, we detail our novel hybrid statistical deep regression model with details of the eGPD and GCNN components provided in Sections~\ref{sec:eGPD} and~\ref{sec:GCNN}, respectively. In Section \ref{sec:Results}, we use our novel framework to jointly model low, moderate, and extreme Australian burnt areas, and provide a wildfire hazard assessment of four population-dense communities: Sydney, Tasmania, Perth, and Melbourne. We identify spatial trends in the frequency and severity of wildfire burnt areas across Australia and these communities, and highlight spatial variation in temporal trends of the wildfire distribution across the domain. We also identify important predictors that drive the frequency and severity of Australian wildfires. In Section \ref{sec:Conclusions}, we conclude the paper with some discussion on future research directions. 


\section{Australian burnt area data} %
\label{sec:DataDescr}
We begin by describing the construction of our novel burnt area data. We first collect burnt area boundaries from the data portal of each Australian state government and combine them into a single database which consists of the boundaries of area burnt by individual wildfires, from June 1999 to December 2018, inclusive \citep[see][]{bush_data}. Boundaries of events in the Northern Territory are of relatively poor quality (i.e., low resolution) or sparsely available, and so we choose to exclude the events in this region from our dataset. We assign, to each boundary, the month of the ignition of its corresponding wildfire. For cases where the ignition date is missing, we assign, in its place, the date of wildfire extinguishment, and assume that this is the same as the missing ignition month. In the very rare case where both ignition and extinguishment dates are missing (1,100 cases out of 241,801), the wildfire events are removed.

For our newly constructed database of wildfire boundaries, we aggregate the burnt areas into monthly values and onto a temporally-consistent set of spatial reference units that are defined by Level~1 Australian Statistical Geography Standard (ASGS) boundaries. These boundaries are used by the Australian Bureau of Statistics to form Statistical Areas Level 1 (SA1) regions and perform population census \citep{australian2011australian}. In most risk reduction activities, local governments have limited jurisdiction outside of their governmental units. By using Level~1 units of ASGS boundaries, we ensure that the considered regions fall well within the smallest governmental units, thus allowing local agencies to use the results of our analysis to inform actionable mitigation policies. In the ``greater'' city areas (i.e., Sydney, Melbourne, Brisbane, Perth, Darwin, Hobart, and Adelaide), where governmental jurisdiction is city-wide, we instead construct polygons using the higher-resolution Level~2 ASGS boundaries. In total, we obtain 7,901 artificially constructed and temporally-consistent polygons that encompass mainland Australia and Tasmania; 311 of these polygons form the Northern Territory. With these reference units, we perform a spatial overlay of the wildfire burnt area polygons and calculate the total burnt area (BA; km$^2$), due to wildfires, within the reference unit in each month.

\begin{figure}[t!]
\centering
\begin{tabular}{cc}
  \includegraphics[width=.48\linewidth]{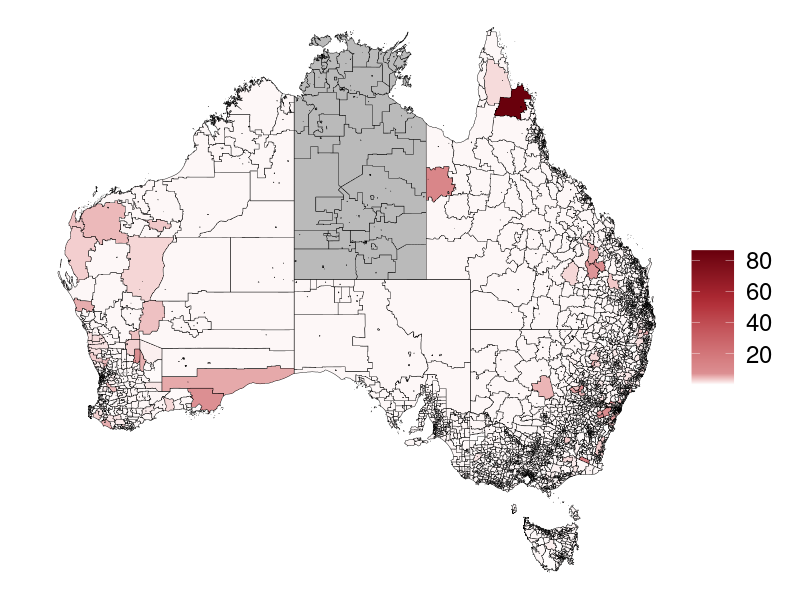} & 
  \includegraphics[width=.48\linewidth]{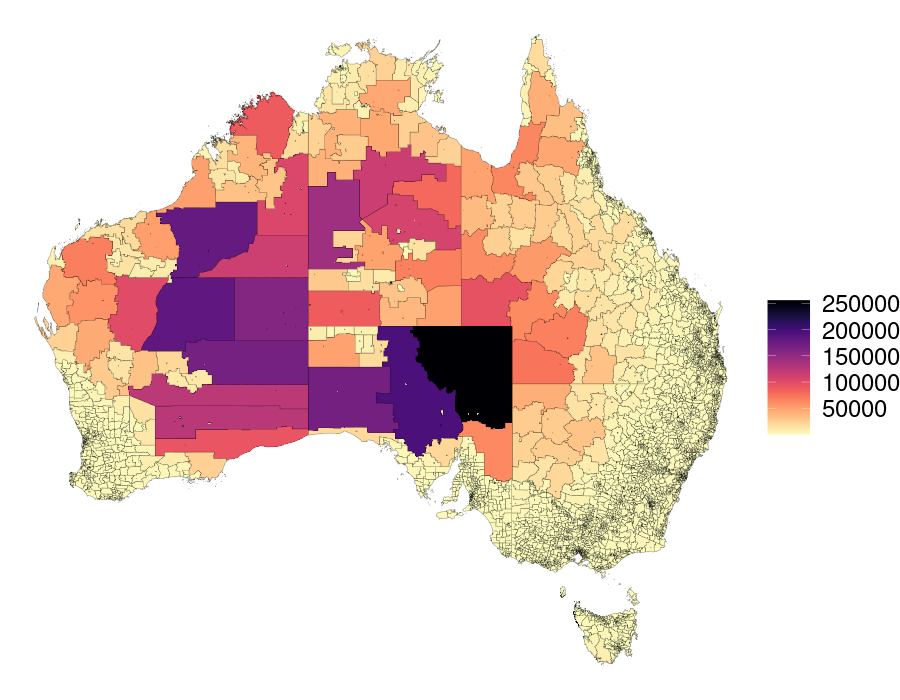} \\
 \includegraphics[width=.48\linewidth]{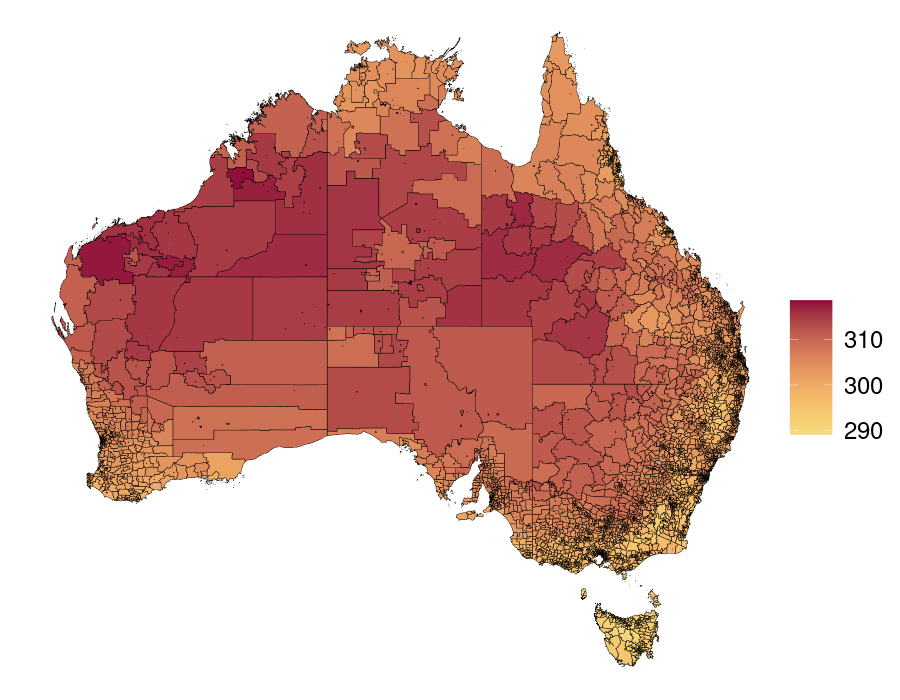} &   \includegraphics[width=.48\linewidth]{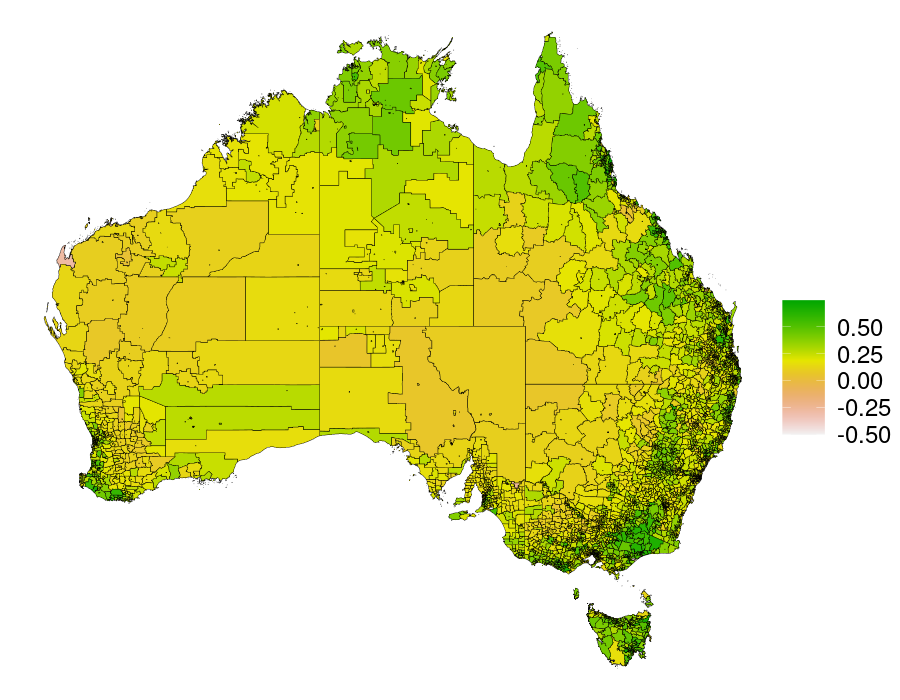} \\
\end{tabular}
\caption{Maps of the monthly radial wildfire spread [$\sqrt{\mbox{BA}}$; km] (top-left), polygon area [km$^2$] (top-right), monthly average air temperature [K] (bottom-left), and NDVI [unitless] (bottom-right) for January 2002. Grey regions in the top-left panel are Statistical Areas Level 1 within the Northern Territory where the response, BA, is not available.}
\label{fig:Observed_Climate}
\end{figure}

To model wildfire occurrence and spread, we incorporate meaningful covariates that play a key role in wildfire dynamics, for example, land-cover types and meteorological conditions \citep{fusco2019invasive, nadeem2020mesoscale}. For the former, we use the monthly average Normalized Difference Vegetation Index (NDVI), which gives a measure of the types of biomass available for ignition in a spatial region (see Figure~\ref{fig:Observed_Climate}). NDVI is a satellite-derived index that reflects the ``greenness" or density of vegetation. Higher NDVI values indicate denser and healthier vegetation, while lower NDVI values suggest sparse or senescing (deteriorating) vegetation. For the meteorological conditions, we include five variables: 2m air temperature (K), precipitation (m), evaporation (m), and both northerly and easterly wind speed components (m/s). For each of these five variables, we use both its monthly average and monthly maximum. Descriptors of the domain topography (i.e., a polygons' average slope ($^{\circ}$) and aspect ($^{\circ}$)) are also available. The monthly covariates we consider are aggregated onto the same spatial reference units as the response variable. Spatial overlay is performed using covariate images obtained from different sources. The static topographical properties (slope and aspect) are derived from the Shuttle Radar Topography Mission digital elevation model \citep{farr2000shuttle}. For the dynamic predictors, monthly averages and maxima of the meteorological variables are obtained from the same hourly values used to derive the ERA-5 monthly land averages \citep{munoz2019era5} and the NDVI is taken from the Tier-1 orthorectified Landsat-7 scenes converted to the top of atmosphere reflectance \citep{chander2009summary}.  Note that, whilst the response data are missing for the Northern Territory, all covariates are available for polygons located within these regions (and can thus be used to make predictions with our model). Monthly aggregated BA and the covariates, are publically available at \url{https://github.com/Jbrich95/pinnEV}, and the individual wildfire boundaries are available upon request.


We focus on modeling the monthly burnt area, which is a measure of the hazard exhibited by a region experiencing wildfires. Following \cite{richards2022unifying}, we model the square-root $\sqrt{\mbox{BA}}$, which provides a measure of the radial spread attributed to wildfires and which leads to improvements in model fits. Figure~\ref{fig:Observed_Climate} illustrates the monthly $\sqrt{\mbox{BA}}$, 2m air temperature, and NDVI, for January 2002, as well as the area of each polygon. We choose to focus on this month in this figure as it experienced particularly devastating wildfires, which occurred in January 2002 and were a part of a larger series of fires that burned across multiple states, including New South Wales, Victoria, and South Australia, and which were exacerbated by prolonged drought, hot temperatures, and strong winds \citep{chafer2004post}. In the span of the following years (2002--2009), Victoria experienced three mega-fires that burnt 40\% of the state’s public land \citep{attiwill2013mega}.
\begin{figure}[t!]
\centering
\begin{tabular}{cc}
\includegraphics[width=.48\linewidth]{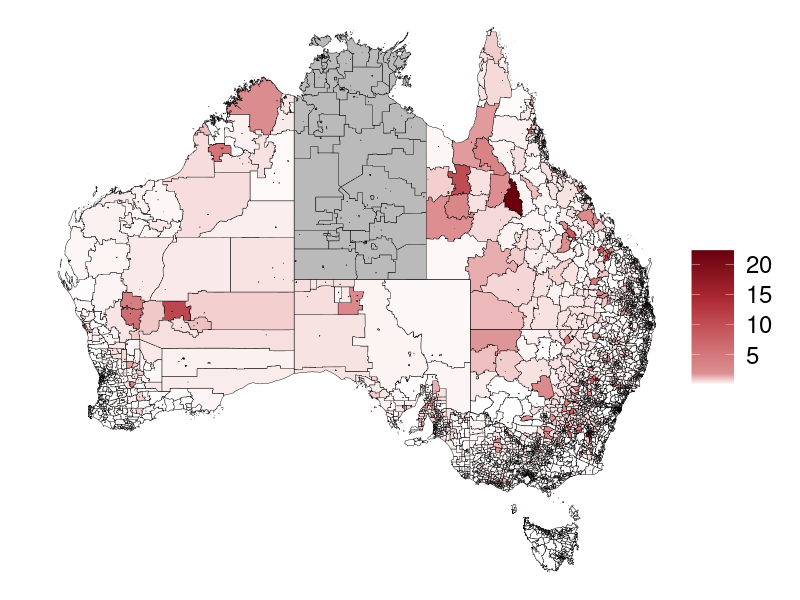} &  \includegraphics[width=.48\linewidth]{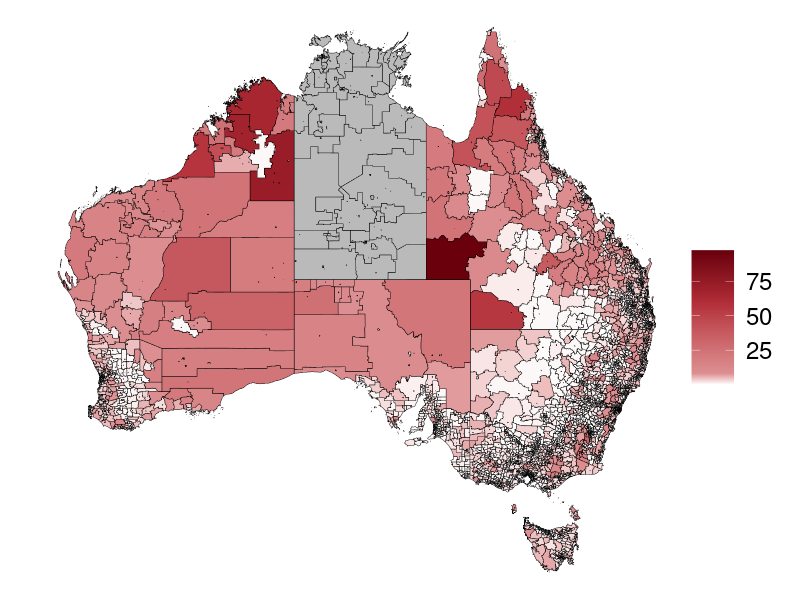} \\
\multicolumn{2}{c}{\includegraphics[width=.98\linewidth]{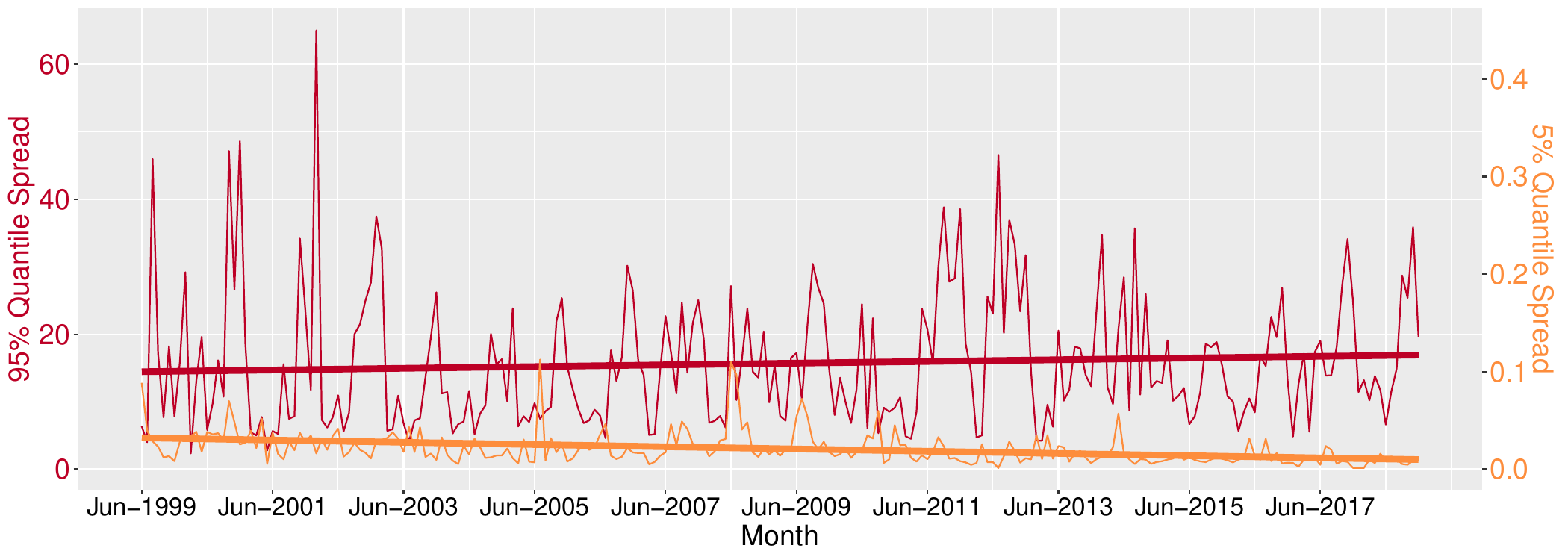}}
\end{tabular}
\caption{Top panels: region-wise $5\%$ (left) and $95\%$ (right) quantiles of the monthly radial wildfire spread [$\sqrt{\mbox{BA}}$; km] pooled across the observation period. Grey regions are Statistical Areas Level 1 within the Northern Territory where the response, BA, is not available. Bottom: time series of $5\%$ (orange) and $95\%$ (red) quantiles of $\sqrt{\mbox{BA}}$ pooled across the entire spatial domain. Straight lines denote estimated least-squares linear trends.}
\label{fig:TS_Spread}
\end{figure}

Figure~\ref{fig:TS_Spread} presents a climatology of the observed wildfires. The top panels map the region-wise 5\% and 95\% quantiles of non-zero values of monthly radial wildfire spread across the entire observation period. While the 5\% quantile tends to have higher magnitudes nearer the mid-Eastern region of Australia, the 95\% quantile shows larger magnitudes near the southern coastal areas, indicating different spatial patterns in the bulk and tail of the response distribution. The bottom panel of Figure~\ref{fig:TS_Spread} provides a time series of the monthly 5\% and 95\% quantiles of the response pooled over the entire spatial domain. We observe a slight positive trend in the $95\%$ quantile of the response, 
 but a negative trend in the lower tail. An exploratory analysis of the data reveals that the total number of wildfires in Australia has increased from 3,685 to 7,299 between 1999 and 2018, with an average of 538 fires per year during the study period. The southern regions of Australia appear to exhibit a higher frequency of wildfires compared to the northern regions, with Victoria and New South Wales having the highest number of fires. The wildfire season also seems to be extending, with a significant increase in the number of fires in January and February. Additionally, the analysis shows that the BA of the wildfires within each polygon varies, with the majority of the fires being small (less than 50km$^2$), and only a small proportion (10\%) of the fires being extremely large (greater than 1,200km$^2$). Motivated by these finding, we include spatial coordinates and time as covariates in our model, allowing us to approximate spatial random effects and account for temporal seasonality and trends when combined, non-linearly, using neural networks.


\section{Methodology}
\label{sec:Methodology}
We present our two-stage deep-learning framework for separate modeling of the occurrence probability of wildfires and the subsequent radial spread of wildfires, conditional on an ignition. 


\subsection{Overview}
Define a spatio-temporal process $\{Y(\bm s,t):\bm s \in \mathcal{S}, t \in \mathcal{T}\}$ indexed by a discrete time domain $\mathcal{T} \subset \mathcal{N}$ and a discrete spatial partitioning $\mathcal{S}=\{\bm s_1, \dots, \bm s_V\}$, which comprises $V$ irregular polygons that satisfy $\cup^V_{i=1} \bm s_i \subseteq \mathbb{R}^2$ and $\cap^V_{i=1} \bm s_i =\emptyset$. For ease of exposition, we hereafter treat each polygon $\bm s \in \mathcal{S}$ as a point location and note that $V=7,901$ in our application.
We then let $\mathbf{X}(\bm s,t)=(X_1(\bm s,t),\dots,X_d(\bm s,t))'$ denote a vector of $d$ predictors available at all space-time locations $(\bm s,t)\in\mathcal{S} \times \mathcal{T}$. We further denote by ${y}(\bm s,t)$ and $\mathbf{x}(\bm s,t)$ the observations of $Y(\bm s,t)$ and $\mathbf{X}(\bm s,t)$, respectively. For our data, the response $Y(\bm s,t)$ is taken to be the square-root of the monthly BA observed over region $\bm s$ in the $t$-th month. The predictor set $\mathbf{X}(\bm s,t)$ is composed of the covariates described in Section~\ref{sec:DataDescr} and, as well as the latitude and longitude coordinates of the centroid of $\bm s$, and $t$-th month and year. We seek to estimate the marginal distribution function of radial spread $Y(\bm s,t)$ conditional on predictors $\mathbf{X}(\bm s,t)=\mathbf{x}(\bm s,t)$, which we denote by $F_{(\bm s,t)}(y)=\Pr\{Y(\bm s,t) \leq y \mid \mathbf{X}(\bm s,t)=\mathbf{x}(\bm s,t)\}$. We assume the form
\begin{equation}
\label{eq:fullmodel}
F_{(\bm s,t)}(y)=\begin{cases} 1-p_0(\bm s,t),\quad &\text{ if } y=0,\\
 1-p_0(\bm s,t)+p_0(\bm s,t)F_{(\bm s,t),+}(y),\quad &\text{ if } y>0,
 \end{cases}
\end{equation}
where $p_0(\bm s,t)=\Pr\{Y(\bm s,t) > 0\; |\; \mathbf{X}(\bm s,t)=\mathbf{x}(\bm s,t)\} \in [0,1]$ is the probability of a wildfire occurrence in region $\bm s$ at time $t$ and where
\begin{equation}
\label{eq:uppermodel}
F_{(\bm s,t),+}(y)=\Pr\{Y(\bm s,t) \leq y \mid Y(\bm s,t) > 0, \mathbf{X}(\bm s,t)=\mathbf{x}(\bm s,t)\}
\end{equation}
is the distribution function of non-zero values of the radial spread $Y(\bm s,t)$, conditional on a wildfire occurrence. We model $p_0(\bm s,t)$ using a deep logistic regression framework. We represent $\log[p_0(\bm s,t)/\{1-p_0(\bm s,t)\}]$ as a function of $\mathbf{x}(\bm s,t)$, which is estimated using the graph convolutional neural network (GCNN) that we describe in Section~\ref{sec:GCNN}. For $F_{(\bm s,t),+}(\cdot)$, we assume the parametric form determined by the extended generalised Pareto distribution (eGPD), which we describe in Section~\ref{sec:eGPD}.

\subsection{Extended generalized Pareto distribution}
\label{sec:eGPD}
Classical asymptotic peaks-over-threshold models utilise the generalised Pareto distribution (GPD) to characterise the upper tail behavior of a random variable \citep[see, e.g.,][]{ pickands1975statistical,davison1990models}; see \cite{Davison.Huser:2015} for a review on extreme value theory. For a random variable $Y$, we assume some high threshold $u$ exists such that $(Y-u)\;|\; (Y > u)$ follows the GPD, denoted $\mbox{GPD}(\sigma_u,\xi)$, for scale and shape parameters $\sigma_u>0$ and $ \xi \in \mathbb{R}$, respectively. Its distribution function is
\begin{align}\label{eq:GPD}
H(y; \sigma_u,\xi)=\begin{cases}
1-(1+\xi y/\sigma_u)_{+}^{-1/\xi},& \xi\neq 0,\\
1-\exp(-y/\sigma_u),&\xi=0,
\end{cases}
\end{align}
for $y>0$ and where $a_{+}=\max(a,0)$. The shape parameter $\xi$ determines the rate of tail decay, with slow power-law decay for $\xi>0$, exponential decay for $\xi=0$, and polynomial decay towards a finite upper bound when $\xi\leq 0$. Although a number of studies have shown that wildfire burnt areas are heavy-tailed, and an appropriate model for them requires $\xi > 0$ \citep{pereira2019statistical,richards2022unifying}, we note that our monthly burnt area data $y(\bm s,t)$ exhibits the natural physical constraint that it cannot be larger than the area of the corresponding SA1/SA2 region. This would suggest that a model with $\xi < 0$ might be more appropriate for our data far in the upper-tail. Following \cite{richards2022insights}, we instead constrain $\xi >0$ throughout as this circumvents numerical issues related to inference of neural network-based regression models with parameter-dependent support \citep[see][]{richards2022unifying}, and this also leads to very good fits with our data all the way from low to high (but not exceedingly high) quantiles (see Section~\ref{sec:results_overview}). Moreover, in Section~\ref{sec:Results}, we estimate $\xi$ to be significantly non-zero, suggesting that $\xi>0$ is a reasonable consistent for our data.

The GPD is a popular choice for modelling the upper-tails of random variables, and has been adopted in the context of modelling wildfires by, for example, \cite{mendes2010spatial}, \cite{turkman2010asymptotic}, \cite{pimont2021prediction} and \cite{richards2022insights}. However, it suffers from three major drawbacks: i) it requires specification of the high exceedance threshold $u$, ii) it does not provide any model for the bulk, $Y\;|\;Y< u$, and iii) the scale parameter $\sigma_u$ depends $u$, making its interpretation difficult \citep{richards2022insights}. Efficient and optimal selection of $u$ is an ongoing problem of discussion in the literature, see, for example, \cite{dupuis1999exceedances} and \cite{deidda2010multiple}; with $u$ too low, the asymptotic arguments that justify the use of the GPD do not hold, and, with $u$ too large, there is not enough data left available for reliable inference. Typically the bulk of the distribution of $Y$ is modelled empirically \citep{davison1990models}. However, in such a case it is difficult to model the impacts of predictors on the distribution of $Y\;|\;Y<u$, which here  corresponds to spread of non-extreme wildfires. Parametric mixture models can be adopted instead \citep{behrens2004bayesian,de2004data,carreau2009hybrid,carreau2011stochastic} , but they impose strong restrictions on the parameter space that would make inference using neural networks computationally problematic. This motivates our use of the extended generalised Pareto distribution (eGPD) as a model for \eqref{eq:uppermodel}, as this can characterise the behaviour of the entirety of $Y$, has asymptotically-justified GPD upper-tails (and lower-tails) and has simple box constraints on the parameter space. \par
 \cite{Papastathopoulos.Tawn:2013}, with extensions by \cite{Naveau.etal:2016}, propose several families of extended GPDs. We adopt the first family \cite[as named by][]{Naveau.etal:2016} and let $Y\sim\mbox{eGPD}(\kappa,\sigma,\xi)$ with distribution function $G(y; \kappa, \sigma, \xi)=\{H(y; \sigma, \xi)\}^\kappa$ for $\kappa >0$, $\sigma>0$, $\xi>0$, and for $H(\cdot)$ in \eqref{eq:GPD}. This model has been previously applied in a regression setting by, for example, \cite{de2021extreme} and \cite{carrer2022distributional}. Here, we take only $\sigma(\bm s,t)$ to be a function of predictors $\mathbf{x}(\bm s,t)$, and leave $\xi$ and $\kappa$ fixed over space and time. Following \cite{richards2022insights}, we use an offset term in the scale function. Defining $a(\bm s) > 0$ as the area of polygon $\bm s$, we let $\sigma(\bm s,t) \propto \sqrt{a(\bm s)}$. We note that different formulations, with heterogeneous $\kappa$ and $\xi$, were also considered in our analysis, but these did not lead to a significant improvement in model fits. The parameter $\sigma$ can be interpreted as a measure of the severity of monthly radial wildfire spread, conditional on at least one wildfire occurrence. We refer to $\sigma$ as the ``conditional spread severity''. Whilst we do not use an offset term for $p_0(\bm s, t)$, we do include $a(\bm s)$ as an extra covariate in $\mathbf{x}(\bm s,t)$ to model wildfire occurences.

\subsection{Deep regression}
\label{sec:GCNN}
We model the occurrence probability $p_0(\bm s,t)$ and conditional spread severity $\sigma(\bm s,t)$ as functions of the predictor set $\{\mathbf{x}(\bm s,t)\}$. Specifically, we let $\mbox{logit}\{p_0(\bm s,t)\}=m_p[\{\mathbf{x}(\bm s,t)\}]$ and $\log\{\sigma(\bm s,t)\}=\log\{\sqrt{a(\bm s)}\}+m_\sigma\{\mathbf{x}(\bm s,t)\}$ for neural networks $m_p$ and $m_\sigma$. In the context of modelling GPD parameters, \cite{rietsch2013network}, \cite{carreau2009hybrid}, \cite{carreau2011stochastic} and \cite{richards2022insights}, illustrate excellent fitting models when using standard densely-connected neural networks (DNNs). \cite{richards2022unifying} advocate the use of convolutional neural networks (CNNs) in place of DNNs, as these can capture spatial structure in the predictors \citep[see review by][]{gu2018recent}. One limitation of CNNs is that they are only applicable to data observed over a regular spatial grid, which is not the case with our data. An alternative family of neural networks are graph neural networks \citep{zhou2020graph}, which are applicable to graphical data, and have already been applied in the context of logistic regression  \citep[see, e.g.,][]{tonks2022forecasting}. A sub-class of graph neural networks are graph convolutional neural networks (GCNNs), which extend the concept of grid-based convolutions to graphs \citep{duvenaud2015convolutional, kipf2016semi}. We represent our spatial data as a graph and investigate the efficacy of GCNNs, relative to DNNs, and now describe both architectures. This architecture has been developed for a general function $m_\star$, where the subscript $\star$ notation can be replaced by $p$ or $\sigma$ as above.\par
A general feed-forward neural network with $J$ hidden layers, each of width $n_j$, can be written as an iterative composition of functions. For the $i$-th node of the $j$-th layer, with $i=1,\dots,n_j$, define a function $m_{j,i}(\bm s^*,t)$, $\bm s^* \in \mathcal{S},$ and the vector of outputs from layer $j$ by $\mathbf{m}_j(\bm s^*,t)=(m_{j,1}(\bm s^*,t),\dots,m_{j,n_j}(\bm s^*,t))'$, with the convention that $n_0=d$ is the number of predictors and $m_{0,i}(\bm s^*,t):=x_i(\bm s^*,t)$ for $i=1,\dots,d$; note that the $n_0$-th layer output is simply the input data $\mathbf{x}(\bm s^*,t)$. The output of the final layer, at region $\bm s^* \in \mathcal{S}$ and time $t$, is
\[
m_{\star}\{\mathbf{x}(\bm s^*,t)\}=(\mathbf{m}_{J}(\bm s^*,t))'\mathbf{w}^{(J+1)}+b^{(J+1)},
\]
for estimable weights and biases $\mathbf{w}^{(J+1)}\in \mathbb{R}^{n_J}$ and $b^{(J+1)}\in\mathbb{R}$, respectively. The output at the $i$-th node of the $j$-th layer, for $j=1,\dots,J$ and $i=1,\dots,n_j$, is
\begin{equation}
\label{layer_eq}
m_{j,i}(\bm s^*,t)=a_j\left\{g_{j,i}(\bm s^*,\{\mathbf{m}_{j-1}(\bm s,t):\bm s \in \mathcal{S}\})+b^{(j,i)}\right\},
\end{equation}
for an estimable bias $b^{(j,i)}\in\mathbb{R}$, fixed activation function $a_j(\cdot)$, and parametric function $g_{j,i}(\cdot,\cdot)$; note that the latter is a function of both the region $\bm s^*$ and $\{\mathbf{m}_{j-1}(\bm s,t):\bm s \in \mathcal{S}\}$, that is, the set of the previous layers features evaluated across the entire spatial domain. Following advocacy by \cite{glorot2011deep}, we take the rectified linear unit (Relu) activation function, that is, $a_j(x)=\max\{0,x\}$, for all $j=1,\dots,J$. If our neural network is a DNN, then we set $g_{j,i}(\cdot,\cdot),$ for all $i=1,\dots,n_j,j=1,\dots,J$, to be 
\begin{equation}
\label{layer_dnneq}
g_{j,i}(\bm s^*,\{\mathbf{m}_{j-1}(\bm s,t):\bm s \in \mathcal{S}\})=(\mathbf{m}_{j-1}( \bm s^*,t))'\mathbf{w}^{(j,i)},
\end{equation}
for weights $\mathbf{w}^{(j,i)}\in \mathbb{R}^{n_{j-1}}$; note that this is a function of $\bm s^*$ only and no other $\bm s \in \mathcal{S}$.

Graph neural networks learn representations of features on a graph $\mathcal{G}=(\mathcal{V},\mathcal{E})$, where $\mathcal{V}$ and $\mathcal{E}$ are a set of $V$ vertices (numbered $1$ to $V$) and corresponding edges, respectively. We represent our spatial domain $\mathcal{S}$ as a graph whereby the centroid of each polygon $\bm s\in\mathcal{S}$ corresponds to a vertex and there exists a (possibly zero) weighted edge between all pairs of regions, $\mathbf{s}_i\neq \mathbf{s}_j$, such that $V:=|\mathcal{V}|=|\mathcal{S}|$. Each spatial polygon, $\bm s_1,\dots,\bm s_V\in \mathcal{S}$, has a corresponding vertex indexed by $v_{\bm s}\in\{1,\dots,V\}$. The graphical structure can be encoded within a symmetric adjacency matrix $A \in \mathbb{R}^{V \times V}$, where each element $A[ij] \geq 0$ denotes the strength of connection between region $\bm s_i$ and $\bm s_j$. We define a graph without self-loops, such that the diagonal elements of $A$ are exactly zero. A typical graph convolutional neural network \citep{kipf2016semi} can be constructed by setting $g_{j,i}(\cdot, \cdot)$ in \eqref{layer_eq}, for all $i=1,\dots,n_j,j=1,\dots,J$, 
to
\begin{equation}
\label{layer_grapheq}
g_{j,i}(\bm s^*,\{\mathbf{m}_{j-1}(\bm s,t):\bm s \in \mathcal{S}\}) =\sum_{\bm s \in \mathcal{S}} \tilde{A}[v_{\bm s^*},v_{\bm s}](\mathbf{m}_{j-1}(\bm s,t))'\mathbf{w}^{(j,i)},
\end{equation}
for estimable weights $\mathbf{w}^{(j,i)}\in \mathbb{R}^{n_{j-1}}$ and where $\tilde{A}=D^{-1/2}AD^{-1/2}$ is the normalised adjacency matrix with $D=\mbox{diag}(\delta_1,\cdots,\delta_v)$ a diagonal matrix with diagonal elements $\delta_i$ set to the degree of the $i$-th vertex (i.e., the sum of elements in the $i$-th row of $A$). Unlike layers in a DNN, the GCNN layer in \eqref{layer_grapheq} pools information across regions in $\mathcal{S}$ that are neighbours of $\bm s^*$, with their contribution to the output determined by the weights in $A$. However, as $A$ (and equivalently $\tilde{A}$) has zeroes along its diagonal, representation \eqref{layer_grapheq} does not exploit information at region $\bm s^*$ directly when updating the features at the same region. We overcome this issue by adding a so-called ``skip connection'', which has been shown to improve training convergence of GCNNs \citep{xu2021optimization}. This is achieved by combining \eqref{layer_dnneq} and \eqref{layer_grapheq}
 and adopting, for all $i=1,\dots,n_j,j=1,\dots,J$, 
\begin{equation}
\label{layer_grapheq_skip}
g_{j,i}(\bm s^*,\{\mathbf{m}_{j-1}(\bm s,t):\bm s \in \mathcal{S}\}) = (\mathbf{m}_{j-1}( \bm s^*,t))'\mathbf{w}_1^{(j,i)} + \sum_{\bm s \in \mathcal{S}} \tilde{A}[v_{\bm s^*},v_{\bm s}](\mathbf{m}_{j-1}(\bm s,t))'\mathbf{w}_2^{(j,i)},
\end{equation}
for weights $\mathbf{w}_1^{(j,i)}\in \mathbb{R}^{n_{j-1}}$, $\mathbf{w}_2^{(j,i)}\in \mathbb{R}^{n_{j-1}}$; we adopt \eqref{layer_grapheq_skip} throughout. Note that equivalence between our GCNN and the DNN is achieved by setting either $A$ to be a matrix of zeroes, that is, having a fully-disconnected graph, or fixing $\mathbf{w}_2^{(j,i)}\in \mathbb{R}^{n_{j-1}}$ to have all-zero entries. \par
In typical applications of graph neural networks, the adjacency matrix $A$ is composed only of binary weights, where $A[ij]=1$ if and only if the $i$-th and $j$-th vertices are connected \citep{wu2020comprehensive}. In our setting, it may be more appropriate to assign non-binary weights to $A$ that reflect the strength of dependence between covariates observed at regions $\bm s \in \mathcal{S}$. In classical geostatistics, this is assumed to be a monotonically decreasing function of pairwise distance $h_{ij}=\|\bm s_i-\bm s_j\|$ for $\bm s_i,\bm s_j \in \mathcal{S}$, where $\|\cdot\|$ is some distance metric \citep{diggle1998model}. In particular, we use the great-circle distance (measured in km) for $\|\cdot\|$. A parameterised distance metric accommodating directionality was also considered, but this did not lead to significant improvements in model fits. 
For the weighted adjacency matrix $A$, we adopt the model 
\begin{equation}
\label{eq:adj}
A[ij]=\begin{cases} \exp\{-(h_{ij}/\lambda)^\alpha\},\quad &\text{ if }h_{i,j} \leq \Delta,i \neq j,\\
0,\quad &\text{ otherwise,}
\end{cases}
\end{equation}
for $\lambda >0$, $\alpha >0$, and $\Delta \geq 0$. The cut-off distance $\Delta$ induces sparsity in $A$; note that if $\Delta=0$, then $A$ has all-zero entries and equivalence between our GCNN and DNN is achieved. Note that $A[ij]=0$ does not explicitly encode into the model that, for sites $\bm s_i$ and $\bm s_j$ such that  $\|\bm s_i - \bm s_j\|>\Delta$,  covariates $\mathbf{X}( \bm s_i, t)$ do not impact the distribution of response $Y( \bm s_j, t)$. Information can pass between non-adjacent nodes in a graph neural network due to a phenomenon called ``message passing'' \citep{zhou2020graph}, which occurs when composing multiple layers of the form given by \eqref{layer_grapheq_skip}.  Due to computational constraints, the adjacency matrix $A$ in \eqref{eq:adj} is not optimised during training of the neural networks characterising $p_0(\bm s, t)$ and $\sigma(\bm s, t)$, and instead its hyperparameters $\lambda$, $\alpha$, and $\Delta$ are manually optimised using cross-validation with a coarse grid-search. To reduce computational time for hyperparameter optimisation, we constrain $\alpha$ to take only two values (i.e., 1 or 2); the case where $\alpha=2$ is equivalent to the thresholded-Gaussian kernel proposed by \cite{shuman2013emerging}.
The use of similarly weighted adjacency matrices in graph neural networks have been exploited previously by, for example, \cite{li2017diffusion}, \cite{yu2017spatio}, and \cite{yao2018deep} in the context of traffic prediction, where their data are point-referenced. We adopt a similar weighting mechanism for our areal data, and illustrate that it facilitates good model fits in Section~\ref{sec:Results}.

\subsection{Inference}
\label{sec:inference}
Inference is conducted by fitting the logistic and the eGPD model \eqref{eq:uppermodel} separately, and combining these to produce estimates of the full distribution in \eqref{eq:fullmodel}. The neural network representations of occurence probability $p_0(\bm s,t)$ and eGPD scale $\sigma(\bm s,t)$ are trained using the the \texttt{R} interface to \texttt{Keras} \citep{kerasforR} and the \texttt{Spektral} Python package \citep{grattarola2021graph}; this procedure is implemented in the \texttt{R} package \texttt{pinnEV} \citep{pinnEV}. We use the Adaptive Moment Estimation (Adam) algorithm \citep{kingma2014adam}, with default parameters and maximal mini-batch size, to minimise the negative log-likelihood associated with $p_0(\bm s, t)$ and the eGPD model. For the eGPD model, we note that the parameters $\kappa$, $\sigma(\bm s,t)$, and $\xi$, are jointly estimated.
\par
For parameter uncertainty quantification, we use a stationary bootstrap \citep{politis1994stationary} to account for temporal dependence, and we set the expected size of spatio-temporal blocks to $k=2$ months. A single bootstrap sample is created by repeating the following until obtaining a sample of length greater than or equal to $|\mathcal{T}|$ months: draw a starting time $t^*\in\mathcal{T}$ uniformly at random and a temporal block size $K\sim\mbox{Geom}(1/k)$; then add the block of observations $\{y(\bm s,t):{\bm s \in \mathcal{S}}, {t\in \{t^*,\dots,t^*+K-1\}}\}$ to the bootstrap sample. If $t^*$ is generated with $t^*+K-1 > |\mathcal{T}|$, instead add $\{y(\bm s,t):\bm s\in\mathcal{S}, t\in \{1,\dots,t^*+K-|\mathcal{T}|-1\}\cup\{t^*,\dots,|\mathcal{T}|\}\}$ and then truncate the sample to have length $|\mathcal{T}|$. For each bootstrap sample, we re-estimate $p_0(\bm s, t)$, the eGPD parameters, and the combined model \eqref{eq:fullmodel}.
\par In order to mitigate over-fitting, we use a robust validation scheme. Each bootstrap sample is randomly partitioned into $80\%$ training and $20\%$ validation data by assigning space-time locations $(\bm s,t)$ to a validation set at random, with equal probability for each $(\bm s, t)$. Neural networks are then trained for a finite number of iterations by minimising the associated negative log-likelihood on the training data, but the final model fit is taken to be that which minimises the negative log-likelihood, estimated on the validation set, across all training iterations. For a single bootstrap sample, the same training and validation sets are used for estimation of $p_0(\bm s, t)$ and the eGPD parameters.

\section{Results}
\label{sec:Results}

\subsection{Overview}
\label{sec:results_overview}
In order to improve numerical stability, prior to training, we standardise each predictor by subtracting and dividing by its marginal mean and standard deviation, respectively. To select the best fitting model, and optimise hyperparameters (i.e., $A$ and Adam learning rate) and neural network architecture, we minimise goodness-of-fit metrics evaluated on the validation data. To assess the classification performance of the occurrence probability model (i.e., the logistic model for $p_0(\bm s,t)$ in \eqref{eq:fullmodel}), we evaluate the area under the receiver operating characteristic curve (AUC). To compare fits of the eGPD model for the conditional radial wildfire spread $Y(\bm s,t)\;|\;Y(\bm s,t)>0$, we utilize the threshold-weighted continuous ranked probability score (twCRPS) \citep{gneiting2011comparing}, which is a proper scoring rule. The twCRPS can be estimated empirically  using a sequence of monotonically increasing thresholds and an appropriate weight function. In our analysis, we use 22 irregularly spaced thresholds ranging from $u_1=0.01$ to $u_{22}=200$, where $u_{22} >\max_{(\bm s,t)\in \mathcal{S}\times\mathcal{T}}\{ y(\bm s,t)\}$ and set the weight function as $r(x)=\tilde{r}(x)/\tilde{r}(u_{19})$, $\tilde{r}(x)=1-(1+(x+1)^2/10)^{-1/4}$, which puts strong focus on calibration in the upper-tail. The estimator of the twCRPS is then
$$\text{twCRPS}=\sum_{(\bm s,t)\in \mathcal{S} \times \mathcal{T}}\sum_{i=1}^{19}r(u_i)[ \mathbbm{1}\{y(\bm s,t)\leq u_i\} -\hat{F}_{(\bm s ,t ),+}(u_i)]^2,$$
where $\hat{F}_{(\bm s,t),+}(\cdot)$ denotes estimates of distribution function \eqref{eq:uppermodel} and $\mathbbm{1}(\cdot)$ is the indicator function. To assess models in the bulk, we also evaluate the regular CRPS (i.e., the twCRPS with constant weight function $r(x)$).

We present model results as pointwise quantiles across 250 bootstrap samples. All neural networks are trained for 3,500 epochs, with the loss evaluated at each iteration using all available observations, that is, with maximal mini-batch size. The optimal architecture for the occurrence probability model $p_0(\bm s, t)$ is a GCNN with $J=3$ hidden layers and $18$ nodes per layer (1483 parameters). For the conditional spread severity $\sigma(\bm s, t)$ from the eGPD model, the optimal architecture is a simpler GCNN with $J=3$, but with 8 nodes per layer (531 parameters), which is unsurprising as less data is available for inference with the eGPD model. For the adjacency matrix $A$ in \eqref{eq:adj}, we found that the optimal range $\lambda$, cut-off distance $\Delta$, and shape $\alpha$ are 650km, 700km, and 2, respectively. The optimal hyperparameters are the same for both the occurrence probability and eGPD model. The median (and $2.5\%$ and $97.5\%$ quantiles) of the bootstrap eGPD shape parameter estimates are $\hat{\kappa}=0.831$ $(0.812,\;0.851)$ and $\hat{\xi}=0.161$ $(0.145, \;0.175)$, suggesting that Australian burnt areas are heavy-tailed. A value of $\hat{\kappa}$ significantly different to one suggests that our model is not equivalent to a regular GPD model. Hence, we have constructed a more flexible model for the bulk and tails of wildfire burnt areas that omits the need for intermediate threshold estimation.

To investigate the increase in model performance when exploiting the GCNN over standard neural networks, we fit the logistic and eGPD models with a DNN in place of the GCNN (see Section~\ref{sec:GCNN}). For fair comparison, we keep the complexity of both models comparable by ensuring that the DNN-based models use the same width and number of layers as the GCNN-based models (as above) and are fitted to the exact same bootstrap samples; we report the median (and $2.5\%$ and $97.5\%$ quantiles) of each diagnostic measure evaluated across all bootstrap samples. For the occurrence probability (logistic) model, the GCNN-based and DNN-based models provide excellent AUC values of  0.919 (0.918, 0.919) and 0.907 (0.906, 0.908), respectively. For the eGPD model, the GCNN-based and DNN-based models provide CRPS values of 16,643 (16,521, 16,770) and 16,658 (16,532, 16,809), respectively, and twCRPS values of 116,488 (114,763, 118,233) and 116,558 (114,464, 118,660), respectively. Whilst the difference in these model diagnostics is fairly moderate overall, we still observe that the use of the GCNN generally yields better performance, especially for the eGPD model. Unreported results also show that the validation loss for the GCNN-based models is consistently lower.

\begin{figure}[t!]
\centering
   \begin{tabular}{cc}
    \includegraphics[width=.42\linewidth]{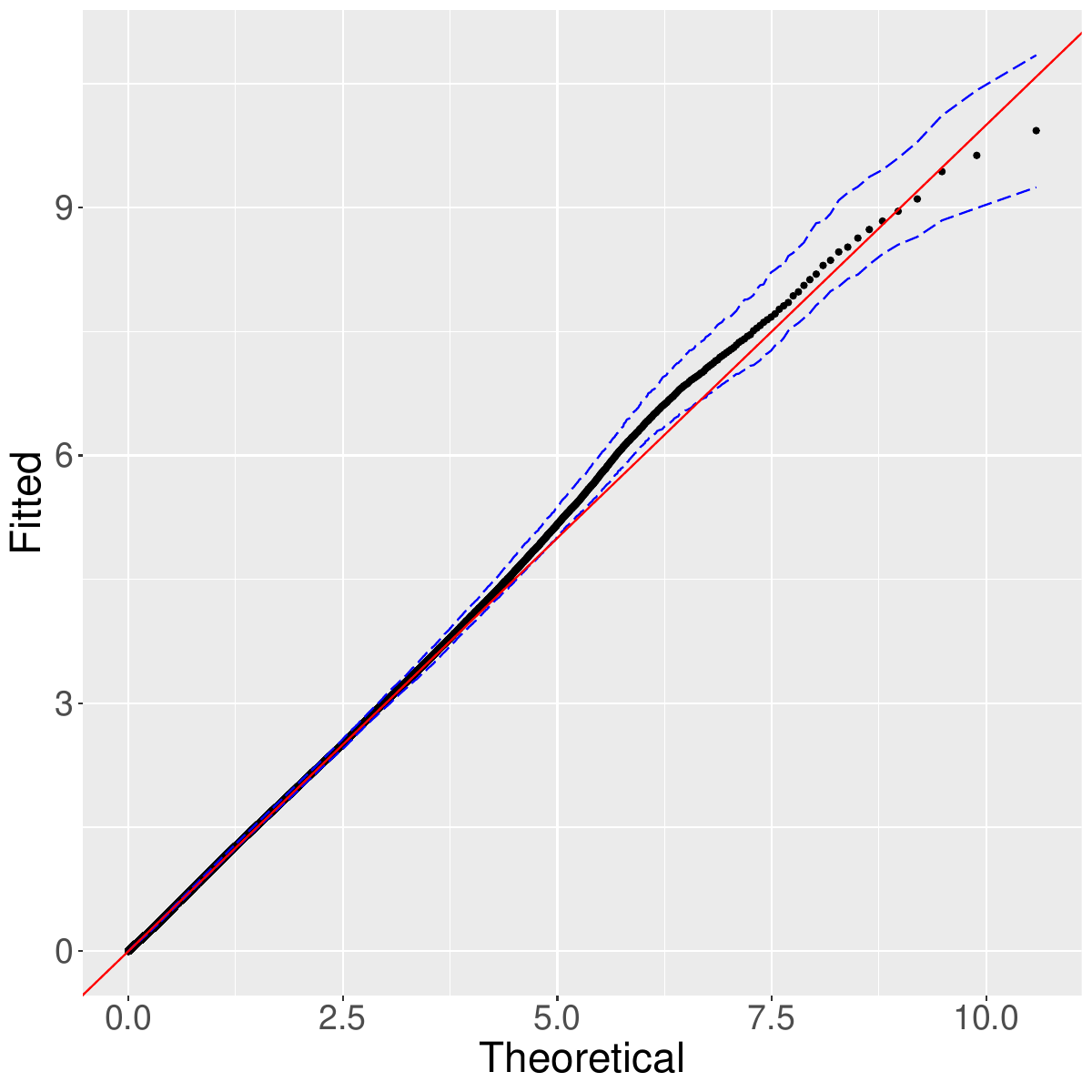}&
     \includegraphics[width=.42\linewidth]{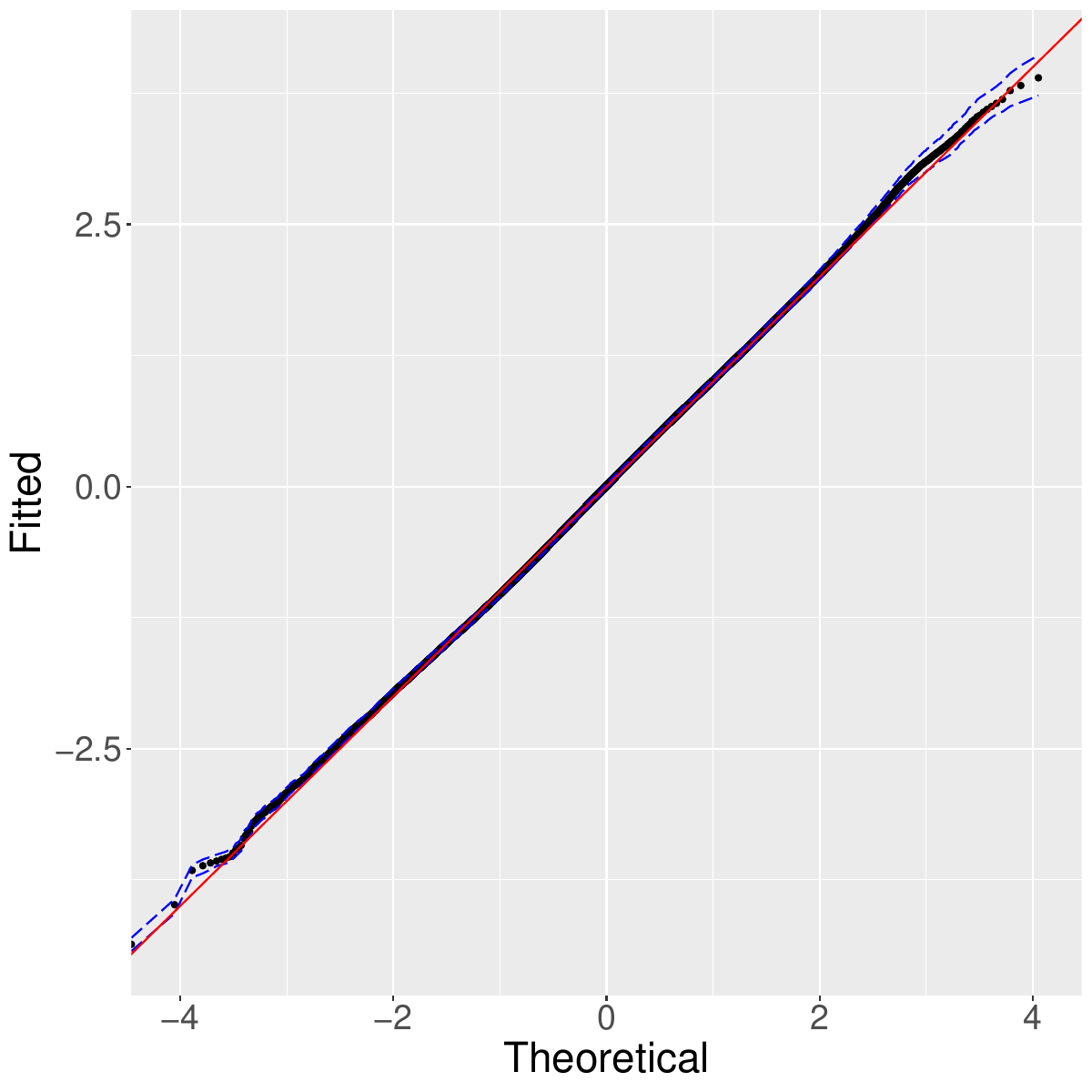}
     \end{tabular}
    \caption{Left: Pooled Q-Q plot for the eGPD model for the monthly radial wildfire spread on standard exponential margins (left), and standard Gaussian margins (right). Data are sampled across the entire spatial domain. 95\% tolerance bounds are given by the blue dashed lines. Black points correspond to observations. }
    \label{fig:QQ-model}
\end{figure}

To visualise the goodness-of-fit for the conditional spread eGPD model, we present a pooled quantile-quantile (Q-Q) plot \citep{heffernan2001extreme} in Figure \ref{fig:QQ-model}. Observations of non-zero radial spread $Y(\bm s,t)\;|\; \{Y(\bm s,t)>0, \bm X(\bm s,t)\}$ are transformed onto standard exponential or Gaussian margins using the estimated eGPD model. We then compare the empirical quantiles of the standardised data against their theoretical counterparts. This process is repeated for each bootstrap sample to build 95\% tolerance intervals. Transforming the data onto standard margins allows us to better assess the fit, with the exponential and Gaussian scales used to better illustrate fits in the upper-tail and bulk, respectively. We observe good model fits in both cases as the estimated 95\% tolerance bands include the diagonal.


\begin{figure}[t!]
\centering
\begin{tabular}{cccc}
  \includegraphics[width=.23\linewidth]{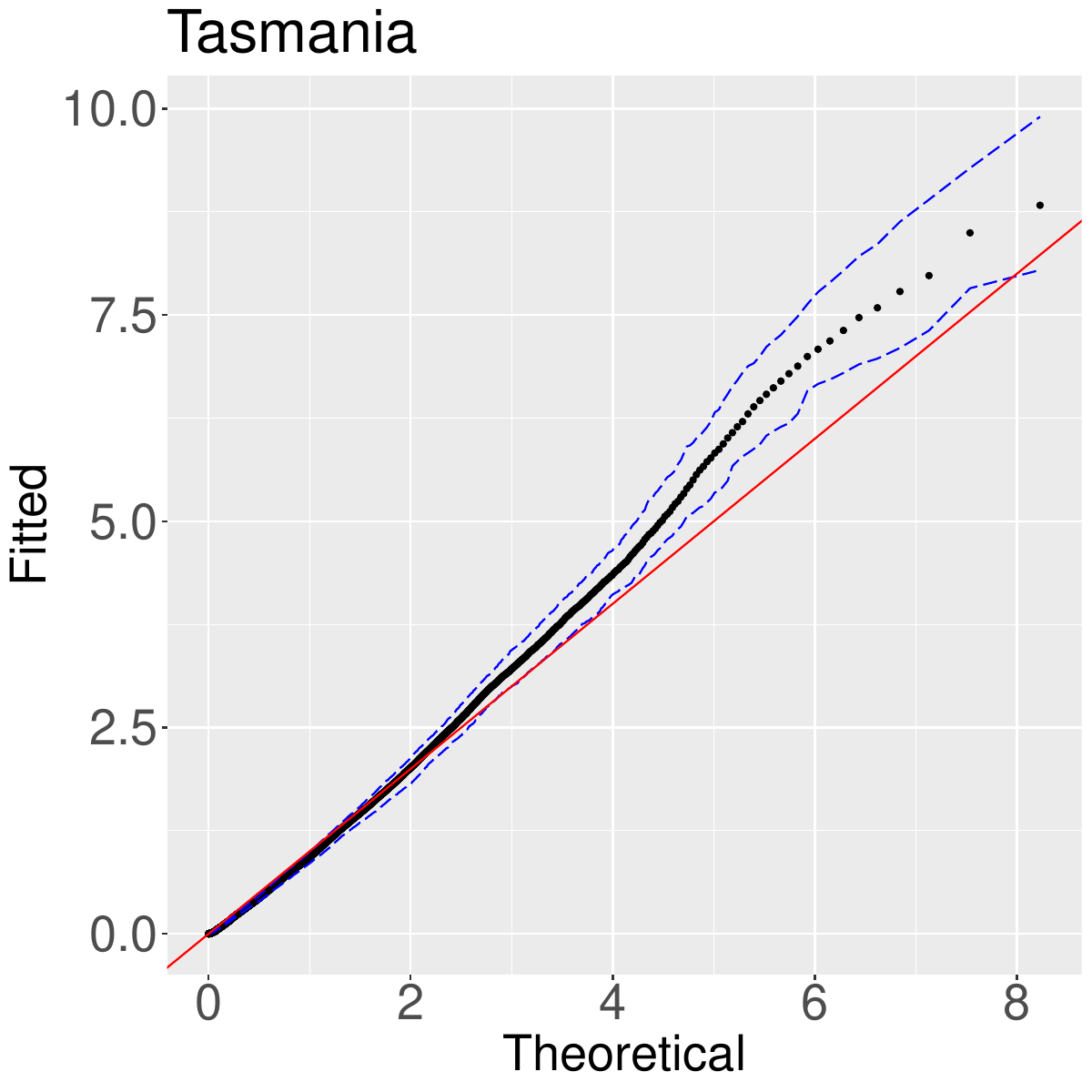} &  \includegraphics[width=.23\linewidth]{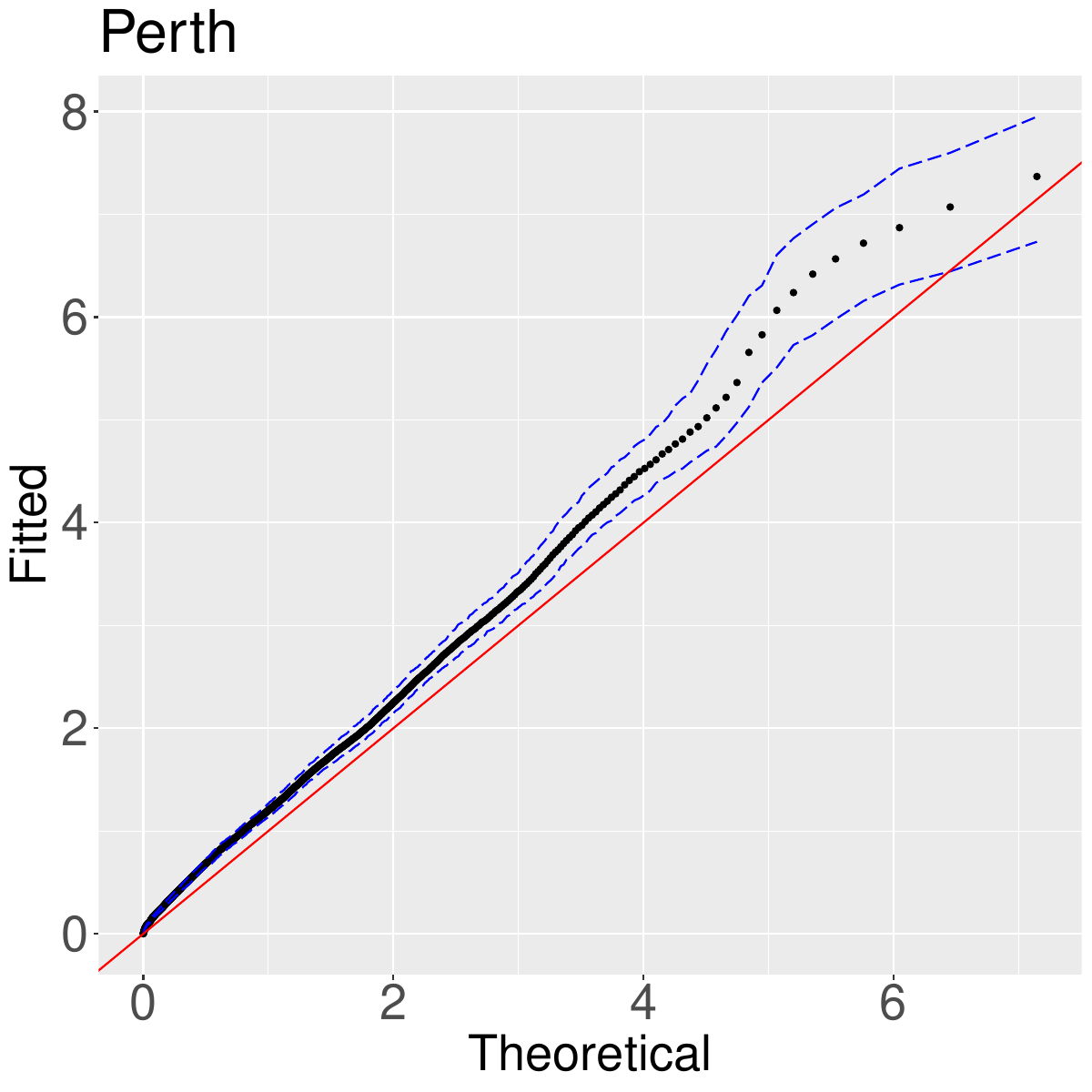} &
 \includegraphics[width=.23\linewidth]{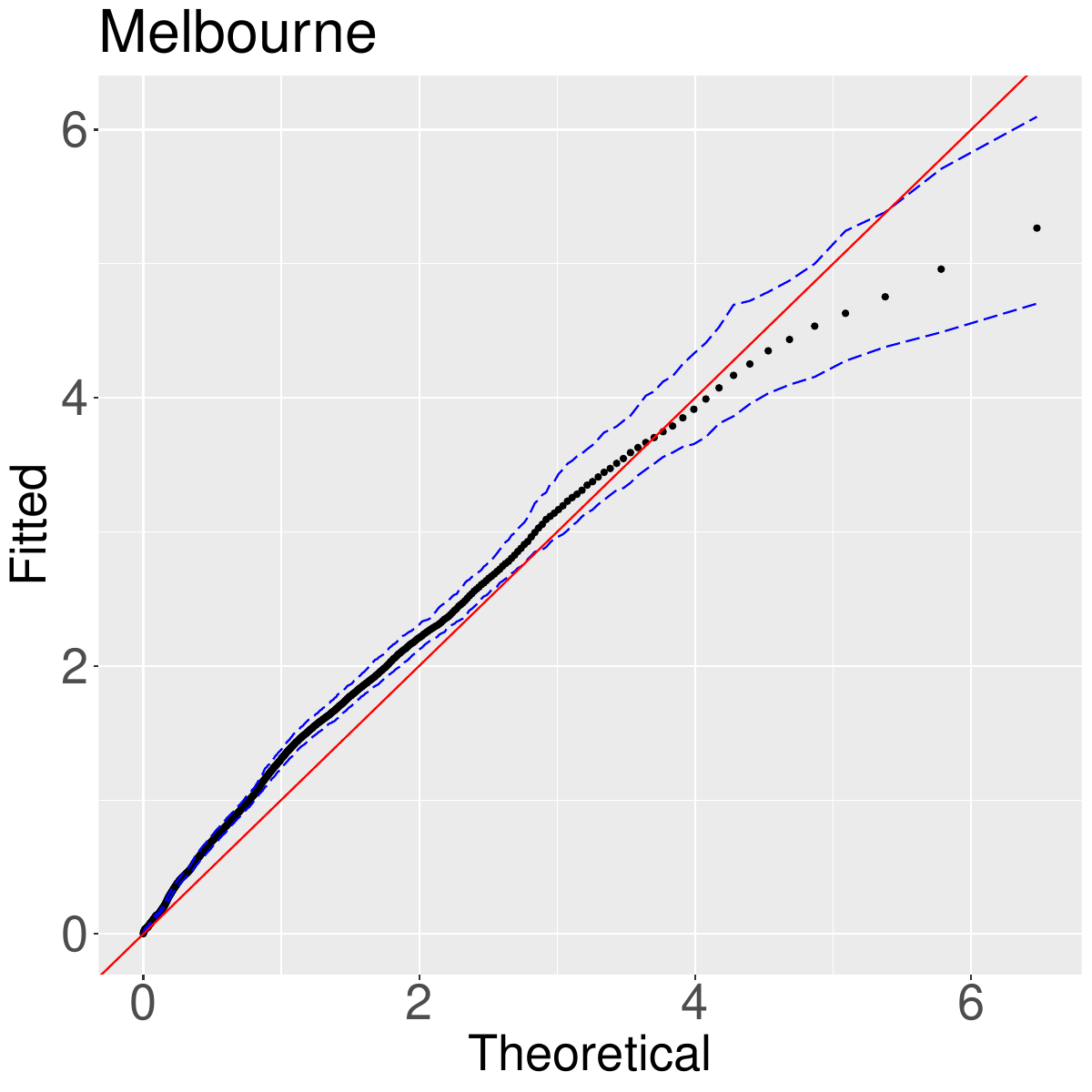} &   \includegraphics[width=.23\linewidth]{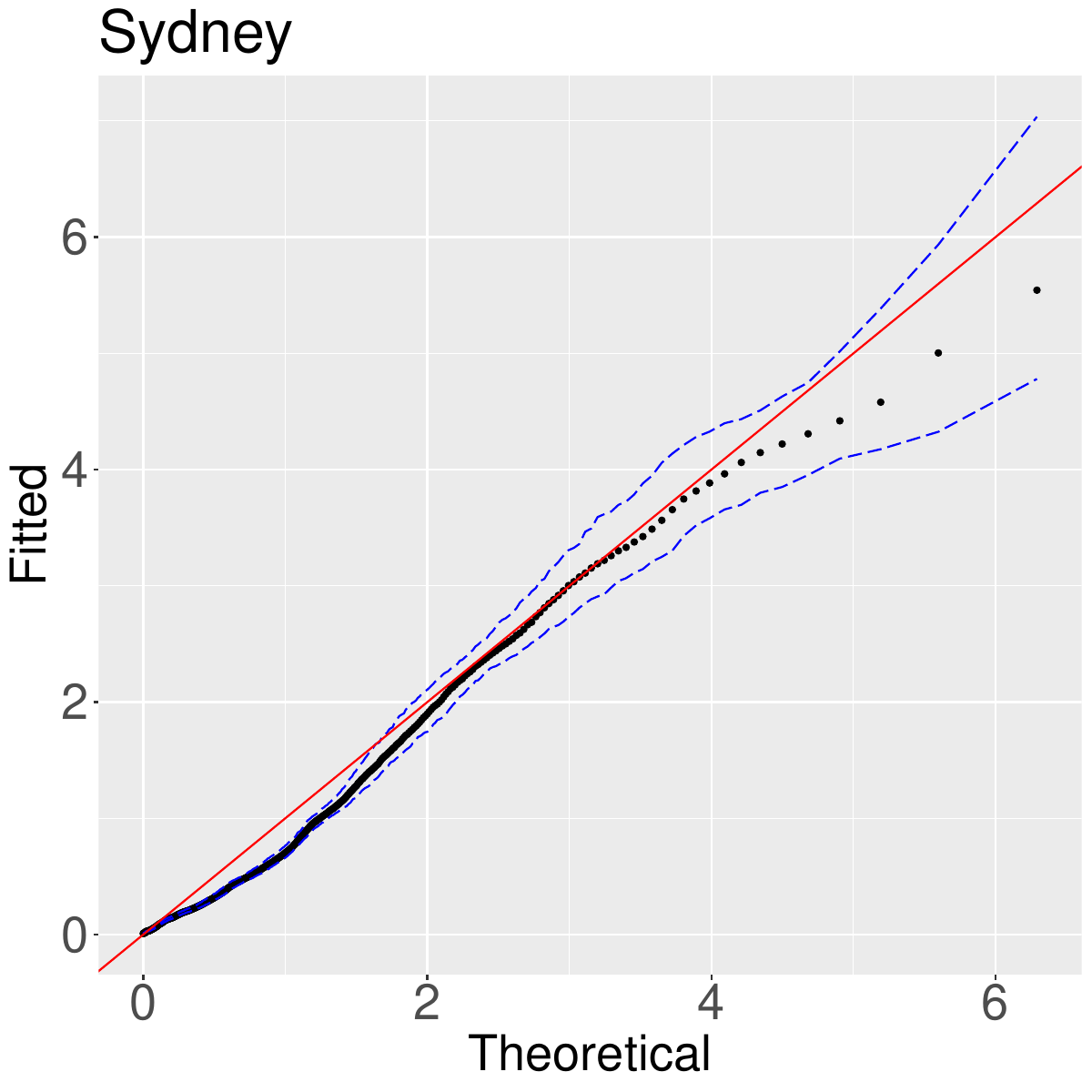} 
\end{tabular}
\caption{Pooled Q-Q plot for the fit of the eGPD model for the monthly radial wildfire spread on standard exponential margins. Data are sampled only in regions corresponding to Tasmania (left), Perth (centre-left), Melbourne (centre-right), and Sydney (right). 95\% tolerance bounds are given by the blue dashed lines. Colored points correspond to observations. }
 \label{fig:QQ-model-Cities}
\end{figure}

Figure~\ref{fig:QQ-model} illustrates good fits of our wildfire spread model globally, that is, the quality of the fit over the entirety of Australia. Figure~\ref{fig:QQ-model-Cities} presents a similar diagnostic (for the eGPD model), albeit using only data sampled within four population-dense communities, namely Tasmania, and areas encompassing Perth, Melbourne, and Sydney. We observe satisfactory local fits of the eGPD model within these population-dense communities, as there is a generally strong alignment between the observed and expected quantiles, though some slight model overestimation of the quantiles is visible for Tasmania and Perth. This suggests that our newly proposed model adequately captures the distribution of the data at local levels quite well, providing further evidence of its efficacy.

\subsection{Wildfire hazard assessment}
\begin{figure}[t!]
\centering
\includegraphics[width=1\linewidth]{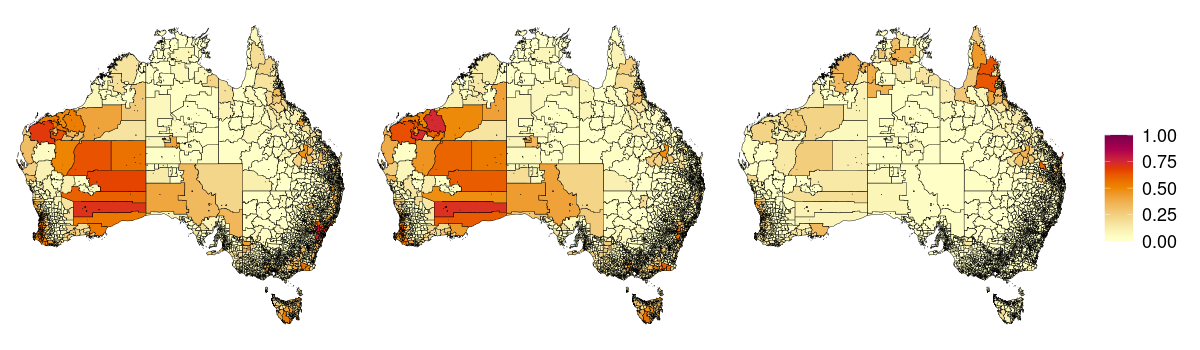} \\
\includegraphics[width=1\linewidth]{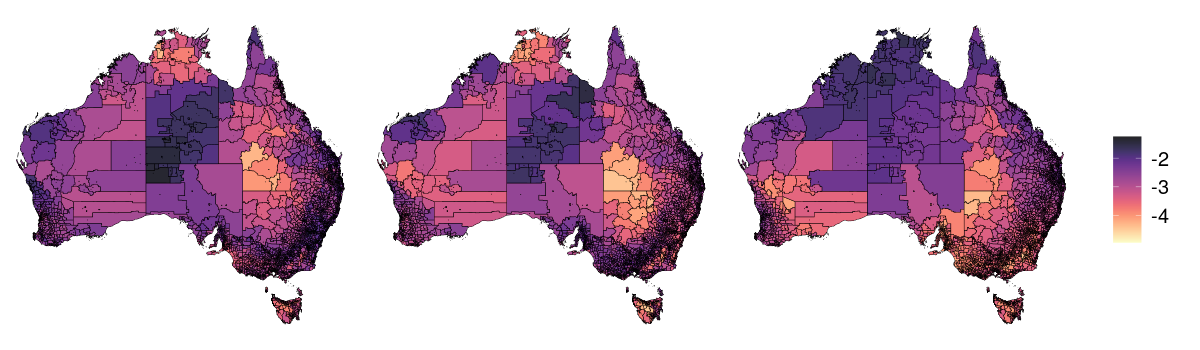} \\
\includegraphics[width=1\linewidth]{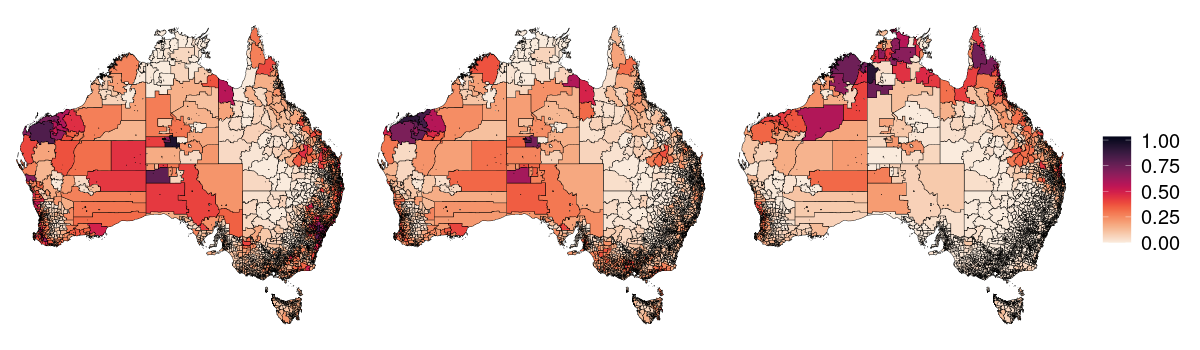}
\caption{Bootstrap median of the estimated probability of wildfire $p_0(\bm s,t)$ [unitless] (first row), relative log-conditional spread severity $\log\{\sigma(\bm s,t)/\sqrt{a(\bm s)}\}$ [unitless] (second row), and relative compound hazard metric CH [unitless] (third row) for January 2002 (left column), January 2005 (center column), and August 2011 (right column).}
\label{fig:estimates}
\end{figure}
We assess the hazard posed by Australian wildfire for three months: January 2002, January 2005, and August 2011. The first two months align with Australia's hot summer, while the last falls within the winter season. The inclusion of months spanning both the hot summer and cooler winter seasons allows us to capture variations in wildfire hazard throughout the year. Whilst no major wildfires occurred in January 2002 or January 2005, significant wildfires ravaged Australia in August 2011, scorching 2.5 million hectares and obliterating 1,800 homes. These fires resulted from a mix of dry weather, strong winds, and lightning strikes. They coincided with a drought, exacerbating the wildfire-prone conditions. 

Figure~\ref{fig:estimates} displays the median estimates (across all bootstrap samples) of the wildfire occurrence probability (i.e., $p_0( \bm s, t)$), the relative log-conditional spread severity $\log\{\sigma(\bm s,t)/\sqrt{a(\bm s)}\}$, and a relative compound hazard measure, denoted by $\textrm{CH}$, which we take to be the 99\% quantile of the square-root of the burnt area divided by polygon area $\sqrt{\mbox{BA}/a(\bm s)}$; note that the latter two measures are standardised by the area of the polygon (i.e., ${a(\bm s)}$), to remove the influence of the area on their value and provide a clearer indication of the areas most hazardous in terms of extreme wildfires. The log-conditional spread severity can be interpreted as a relative measure of wildfire spread severity, given a wildfire occurrence, whilst the compound hazard measure combines information from both the spread and occurrence models. 

For all selected months, we observe generally low occurrence probability estimates across most regions of Australia. In the January months, the locations with higher probability of wildfire occurrence (i.e., $p_0(\bm s, t)$) are generally concentrated in the western and southern regions of Australia, whereas in August 2011 we instead observe the highest estimates in the north east.
Across all selected months, we observe generally lower estimates of the conditional spread severity in the eastern regions of Australia and higher values in the northern areas. Similarly to $p_0(\bm s, t)$, different spatial patterns are observed in January and August; the areas of lowest severity tend to be located in the South for August. We note that regions of high probability of wildfire occurrence do not necessarily correspond to regions of high conditional spread severity, suggesting different spatial trends in the behaviour of these two model components. Maps of the estimated compound hazard do not exhibit the same spatial patterns as those for $p_0(\bm s, t)$ and $\sigma(\bm s, t)$. In January, there is little spatial coherence in the regions most-at-risk of wildfires; in August, the regions that experience the highest compound hazard are generally located along the north coast. 
\begin{figure}[t!]
    \centering
   \includegraphics[width=.985\linewidth]{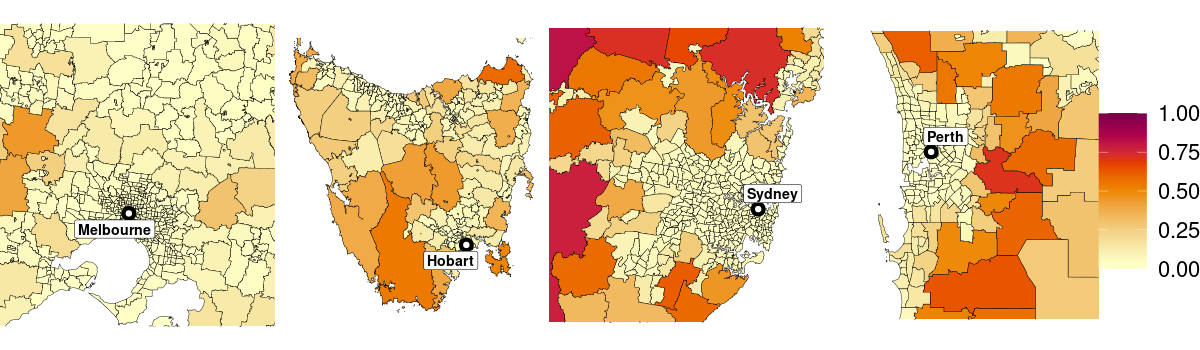}
 \includegraphics[width=.985\linewidth]{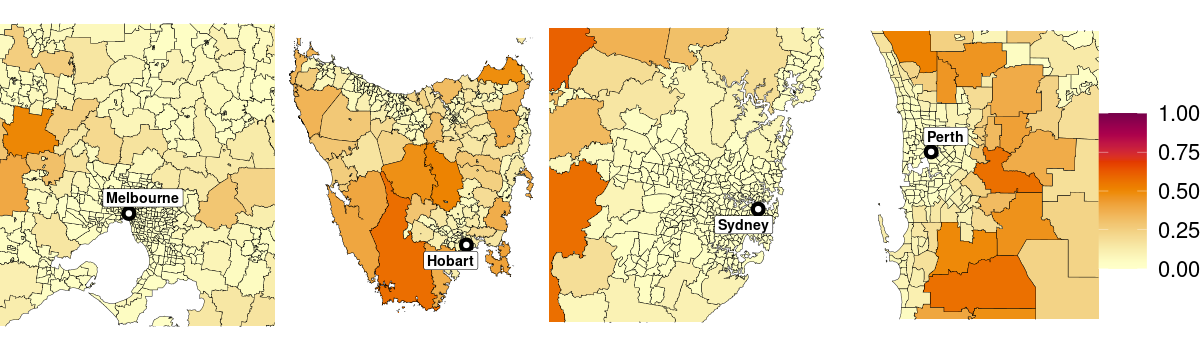}
 \includegraphics[width=.985\linewidth]{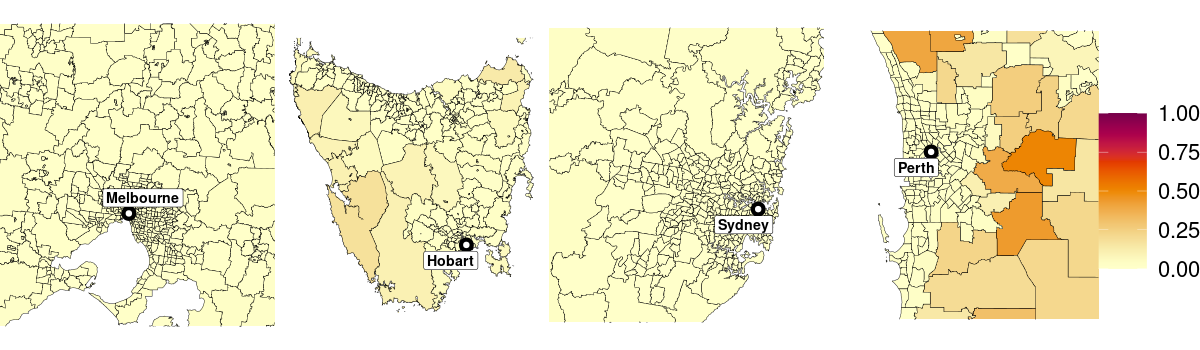}
 \caption{Bootstrap median of the estimated probability of wildfire $p_0(\bm s,t)$ [unitless] for January 2002 (top row), January 2005 (second row), and August 2011  (third row). Columns correspond to spatial domains of Melbourne, Tasmania, Sydney, and Perth (left to right). }
\label{fig:estimates-citiesP0}
\end{figure}
\begin{figure}[t!]
    \centering
   \includegraphics[width=.985\linewidth]{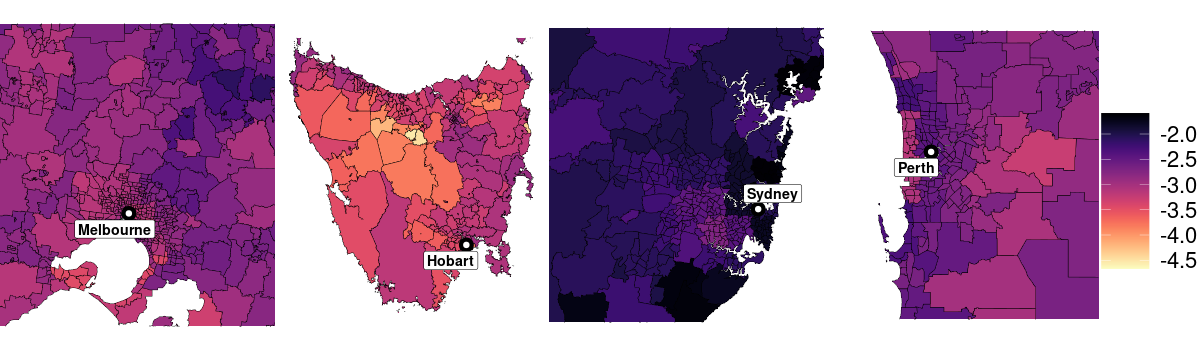}
 \includegraphics[width=.985\linewidth]{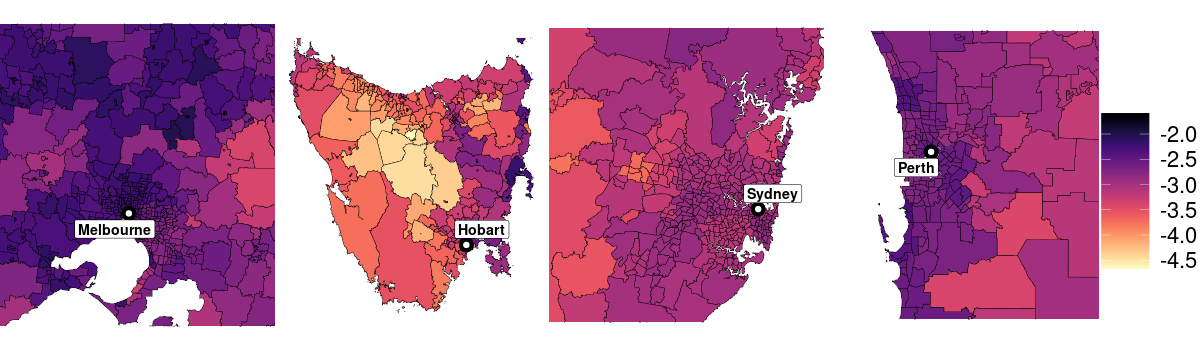}
 \includegraphics[width=.985\linewidth]{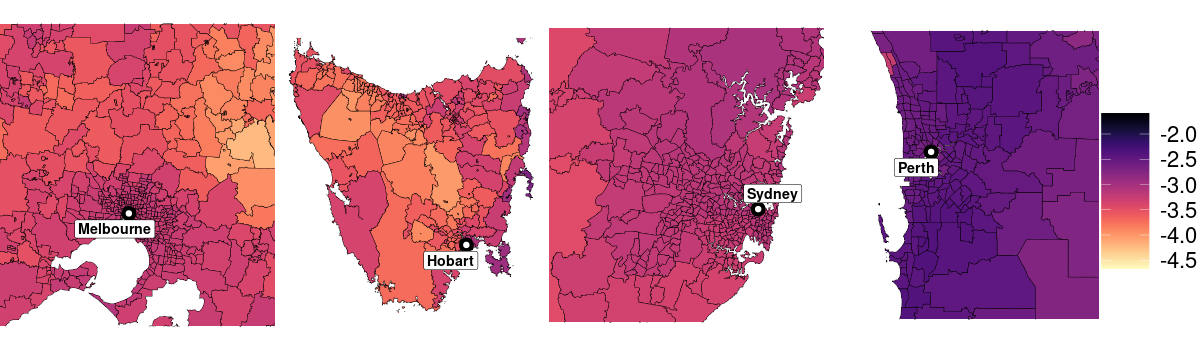}
 \caption{Bootstrap median of the estimated relative log-conditional spread severity $\log\{\sigma(\bm s,t)/\sqrt{a(\bm s)}\}$ [unitless] for January 2002 (top row),  January 2005 (second row), and August 2011 (third row). Columns correspond to spatial domains of Melbourne, Tasmania, Sydney, and Perth (left to right).}
\label{fig:estimates-citiesSig}
\end{figure}

\begin{figure}[t!]
    \centering
   \includegraphics[width=.985\linewidth]{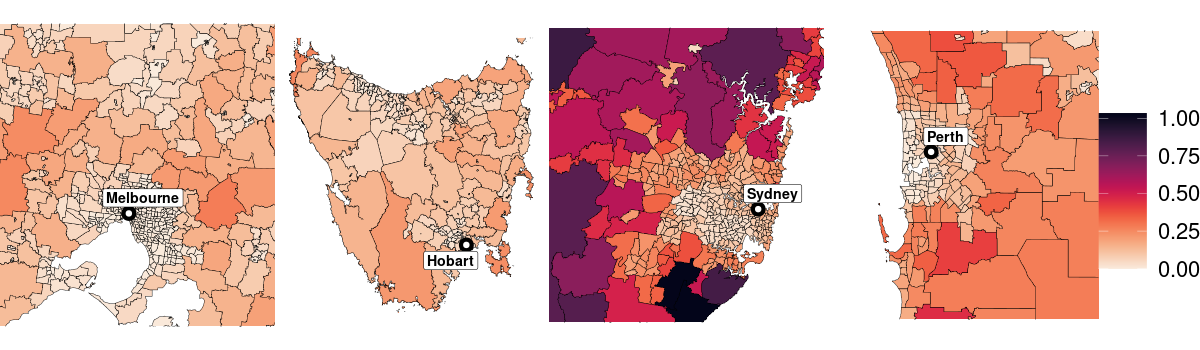}
 \includegraphics[width=.985\linewidth]{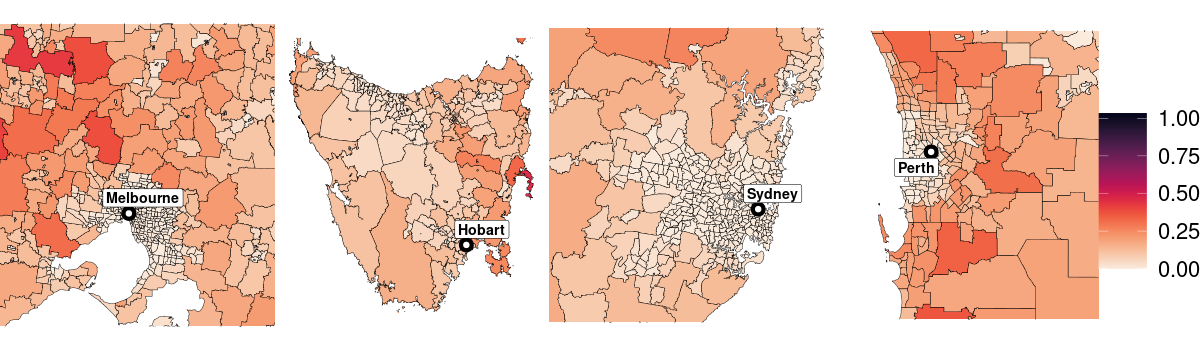}
 \includegraphics[width=.985\linewidth]{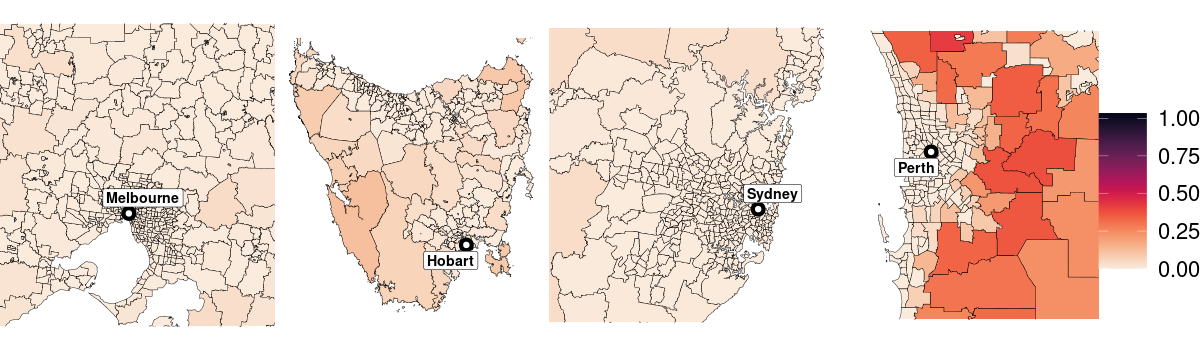}
 \caption{Bootstrap median of the estimated relative compound hazard metric CH [unitless] for January 2002 (top row), January 2005 (second row), and August 2011 (third row). Columns correspond to spatial domains of Melbourne, Tasmania, Sydney, and Perth (left to right).}
\label{fig:estimates-citiesCR}
\end{figure}
Figures \ref{fig:estimates-citiesP0}--\ref{fig:estimates-citiesCR} display estimates of the model summaries, detailed above, for the population-dense communities of Tasmania and the neighborhoods surrounding Perth, Melbourne, and Sydney, for the same three selected months detailed above. During the summer months, a noticeable trend emerges whereby the probability of wildfire occurrence (Figure~\ref{fig:estimates-citiesP0}) is higher in the regions surrounding the major cities. In contrast, the conditional spread severity (Figure~\ref{fig:estimates-citiesSig}) is notably higher within densely populated regions, particularly in Perth, which stands out as the sole area exhibiting any significant hazard during the winter season. We observe generally higher estimates of both the wildfire occurrence probability and conditional spread severity in the hotter January months across all considered communities. Amongst the regions, and across all months, Tasmania exhibits the lowest values of the estimated relative compound hazard metric (see Figure~\ref{fig:estimates-citiesCR}). However, we note that the Southwest National Park in Tasmania, a major national park known for its abundance of burnable material \citep{colhoun1999late}, exhibits the highest hazard within Tasmania. For the other regions, low compound hazard estimates are generally observed in the central, densely populated regions with increasingly large estimates for regions further from the corresponding city's centre.

\begin{figure}[t!]
\includegraphics[width=.98\linewidth]{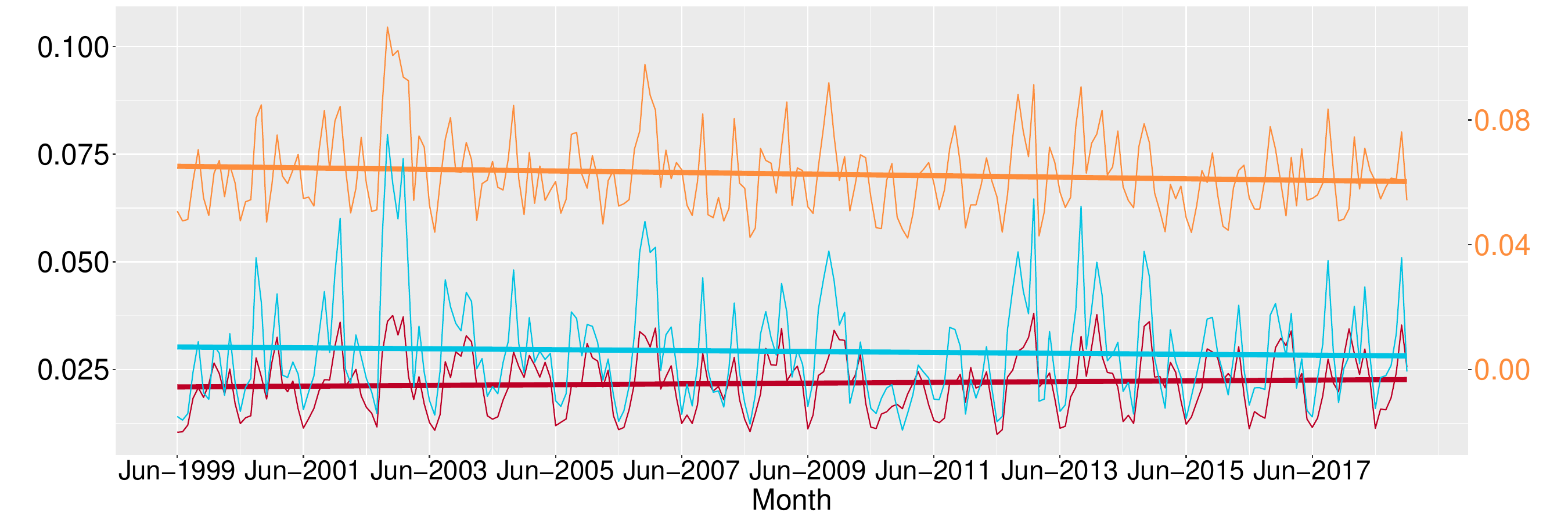}
\caption{Time series of bootstrap median estimates of spatially-averaged hazard metrics: probability of wildfire occurrence $p_0(\bm s, t)$ [unitless] (red), relative conditional spread severity $\sigma(\bm s, t)/\sqrt{a(\bm s)}$ [unitless] (orange), and relative compound hazard metric CH [unitless] (blue). Straight lines denote estimated least-squares linear trends. The left-hand $y$-axis (black) corresponds to $p_0(\bm s, t)$ and CH, whilst the right-hand $y$-axis (orange) corresponds to $\sigma(\bm s, t)/\sqrt{a(\bm s)}$. Metrics are averaged over the entire spatial domain.}
\label{fig:Res:trend}
\end{figure}

\begin{figure}[t!]
\centering
\includegraphics[width=.85\linewidth]{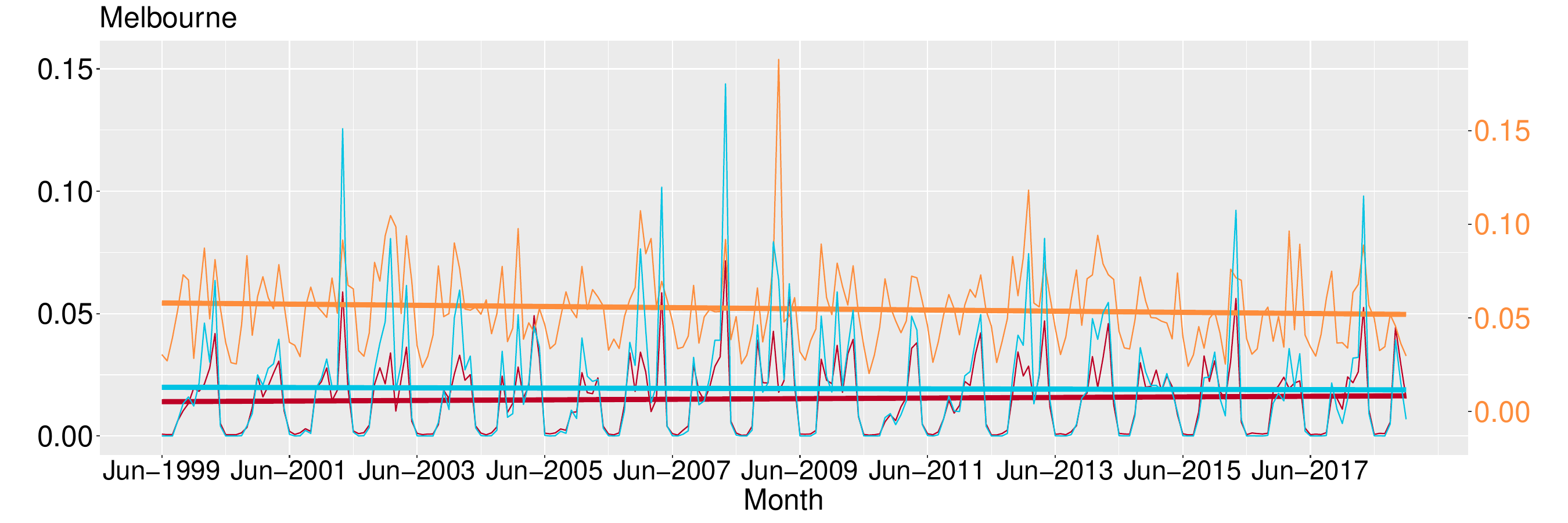}
\includegraphics[width=.85\linewidth]{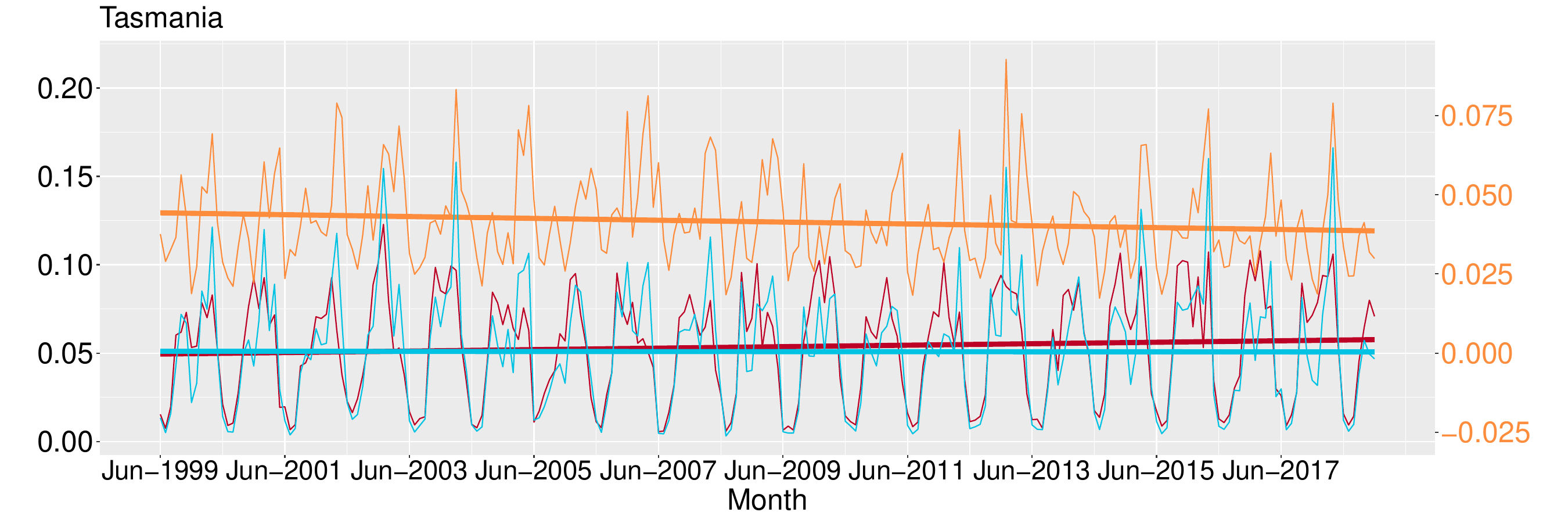}
\includegraphics[width=.85\linewidth]{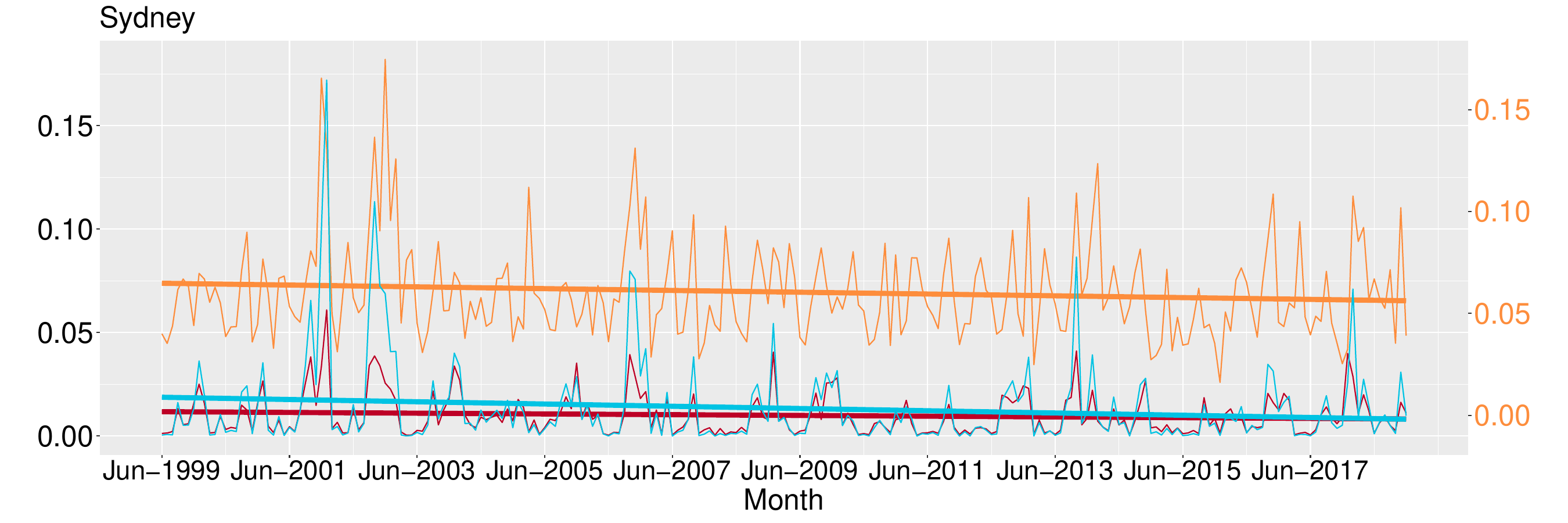}
\includegraphics[width=.85\linewidth]{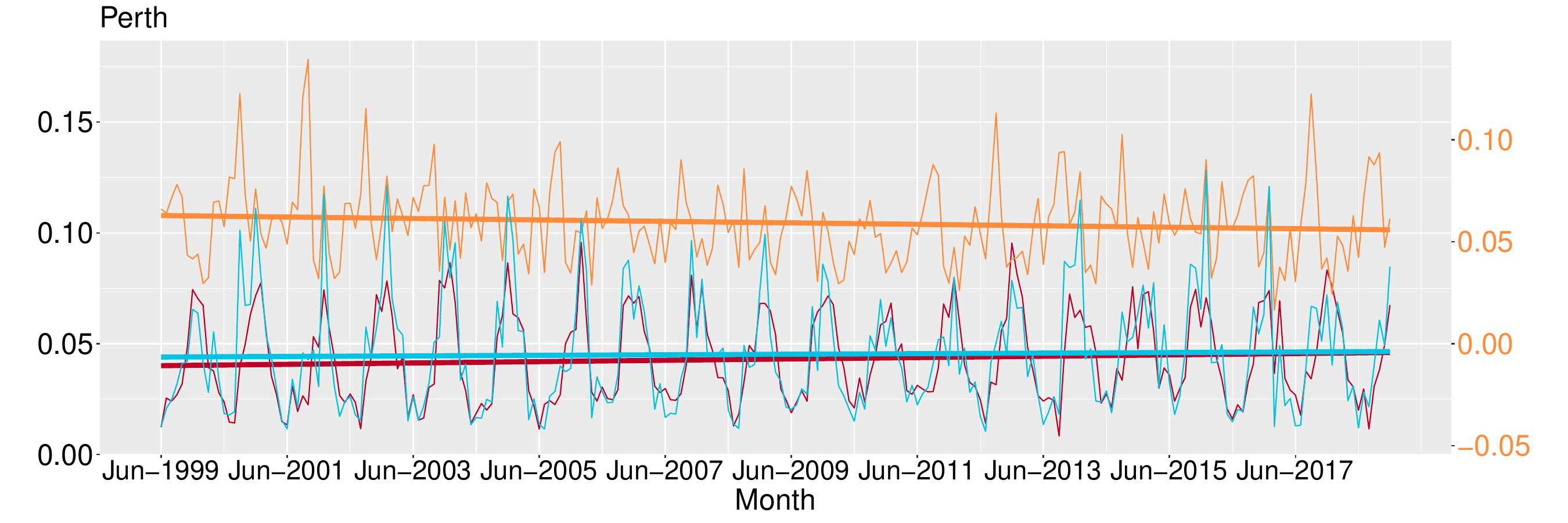}
\caption{Time series of bootstrap median estimates of spatially-averaged hazard metrics: probability of wildfire occurrence $p_0(\bm s, t)$ [unitless] (red), relative conditional spread severity $\sigma(\bm s, t)/\sqrt{a(\bm s)}$ [unitless] (orange), and relative compound hazard metric CH [unitless] (blue). Straight lines denote estimated least-squares linear trends. The left-hand $y$-axis (black) corresponds to $p_0(\bm s, t)$ and CH, whilst the right-hand $y$-axis (orange) corresponds to $\sigma(\bm s, t)/\sqrt{a(\bm s)}$. Metrics are averaged over Melbourne, Tasmania, Sydney, and Perth (top to bottom rows).}
\label{fig:Res:trend_cities}
\end{figure}

To assess temporal changes in the wildfire distribution,  we calculate the spatial mean of each of the three considered hazard metrics ($p_0( \bm s,t)$, $\sigma(\bm s, t)/\sqrt{a(\bm s)}$, and CH) and plot the time series of the median of their bootstrap estimates. Figures~\ref{fig:Res:trend} and \ref{fig:Res:trend_cities} illustrate these time series for averages over the entire spatial domain and the considered population-dense communities (as shown in Figures~\ref{fig:estimates-citiesP0}--\ref{fig:estimates-citiesCR}), respectively. Figure~\ref{fig:Res:trend} illustrates slight positive and negative trends in the frequency and severity (conditional on an ignition), respectively, of wildfires across the time period. However, the resulting trend in the compound hazard metric is slightly negative, which may suggest that increasing wildfire frequency is not generating increased impact due to decreasing severity. Figure~\ref{fig:Res:trend_cities} illustrates that there may be spatial variation in the temporal trends. We observe slight negative trends for the relative conditional spread severity across all four communities, but only Sydney appears to exhibit a negative trend in wildfire frequency. After combining changes in $p_0(\bm s,t)$ and $\sigma(\bm s, t)/\sqrt{a(\bm s)}$ to obtain CH, we observe positive trends for Perth and negative trends for Melbourne and Sydney. In Tasmania, positive trends in  $p_0(\bm s,t)$ and negative trends in $\sigma(\bm s, t)/\sqrt{a(\bm s)}$ appear to cancel each other out as (visually) there is no significant trend in CH.

\subsection{Variable importance}
To rank the importance of individual covariates in the estimation of $p_0(\bm s, t)$ and $\sigma(\bm s, t)$, we use Deep Learning Important Features (DeepLIFT); see \cite{shrikumar2017learning, yuan2022explainability}. This algorithm quantifies the impact of a covariate on the parameter estimates, in a post hoc manner, via contribution scores. Such scores measure the change in the partial gradient of the parameter function at location $(\mathbf{s},t)$ (with respect to the $i$-th covariate at the same space-time location) evaluated for both the observed and a reference set of covariates. DeepLIFT is considered a fast approximation to Shapley values, which have been previously used to interpret deep learning regression models \citep[see, e.g.,][]{lundberg2017unified, dahal2023explainable, wikle2023illustration}.\par
We detail the derivation of the individual contribution scores for the parameter $p_0(\bm s, t)$ only; however, they can be similarly derived for the conditional spread severity $\sigma(\bm s, t)$. Define $g_{(\mathbf{s},t),i}(\mathbf{x}^*_t)$, $i=1,\dots,d,(\mathbf{s},t)\in\mathcal{S}\times\mathcal{T},$ as the scalar partial derivative $\partial p_0(\mathbf{s},t)/\partial {x}_i(\mathbf{s},t)$ evaluated at the observed covariate set $\mathbf{x}_t^*:=\{\mathbf{x}(\bm s,t):\mathbf{s}\in\mathcal{S}\}$ (for details on how $p_0(\mathbf{s},t)$ functionally relates to ${x}_i(\bm s,t)$, see Section~\ref{sec:GCNN}). We then construct some reference set $\mathbf{x}^{(0)}_t$, here taken to be $\mathbf{x}^*_t$ but with all zero entries; recall from Section~\ref{sec:results_overview} that this corresponds to all covariates fixed to their respective mean value. Following \cite{shun2020evaluation}, the contribution score $\mbox{CS}_{i,(\mathbf{s},t)}$ for the effect of predictor ${x}_i(\mathbf{s},t)$ on $p_0(\mathbf{s},t),(\mathbf{s},t)\in\mathcal{S}\times\mathcal{T}$, $i=1,\dots,d,$ is
\[
\mbox{CS}_{i,(\mathbf{s},t)}=x_i(\mathbf{s},t){g_{(\mathbf{s},t),i}(\mathbf{x}^*_t)\over g_{(\mathbf{s},t),i}(\mathbf{x}^*_t)-g_{(\mathbf{s},t),i}(\mathbf{x}^{(0)}_t)}.
\]
We can interpret $\mbox{CS}_{i,(\mathbf{s},t)}$ as the amount of difference-from-mean in $p_0(\mathbf{s},t)$ that can be attributed to the difference-from-mean of $x_i(\mathbf{s},t)$ \citep{shrikumar2017learning}. We estimate the contribution scores for both $p_0(\bm s, t)$ and $\sigma(\bm s, t)$, and for all covariates $i=1,\dots,d$, and all space-time locations $(\mathbf{s},t)\in\mathcal{S}\times\mathcal{T}$. Note that we cannot directly infer the direction of the effect of a covariate $x_i(\mathbf{s},t)$, for a single $(\bm s, t),$ on a model parameter from its score $\mbox{CS}_{i,(\mathbf{s},t)}$ alone, as the value of the latter depends on whether or not $x_i(\mathbf{s},t)$ exceeds its marginal mean. Instead, we pool estimates of $\mbox{CS}_{i,(\mathbf{s},t)}$ over all $(\bm s, t)$. As these estimates span zero, we rank the impact of covariates on the model parameters by the magnitude of the variability in their corresponding estimates.

\begin{figure}[t!]
\centering
    \includegraphics[width=0.96\linewidth]{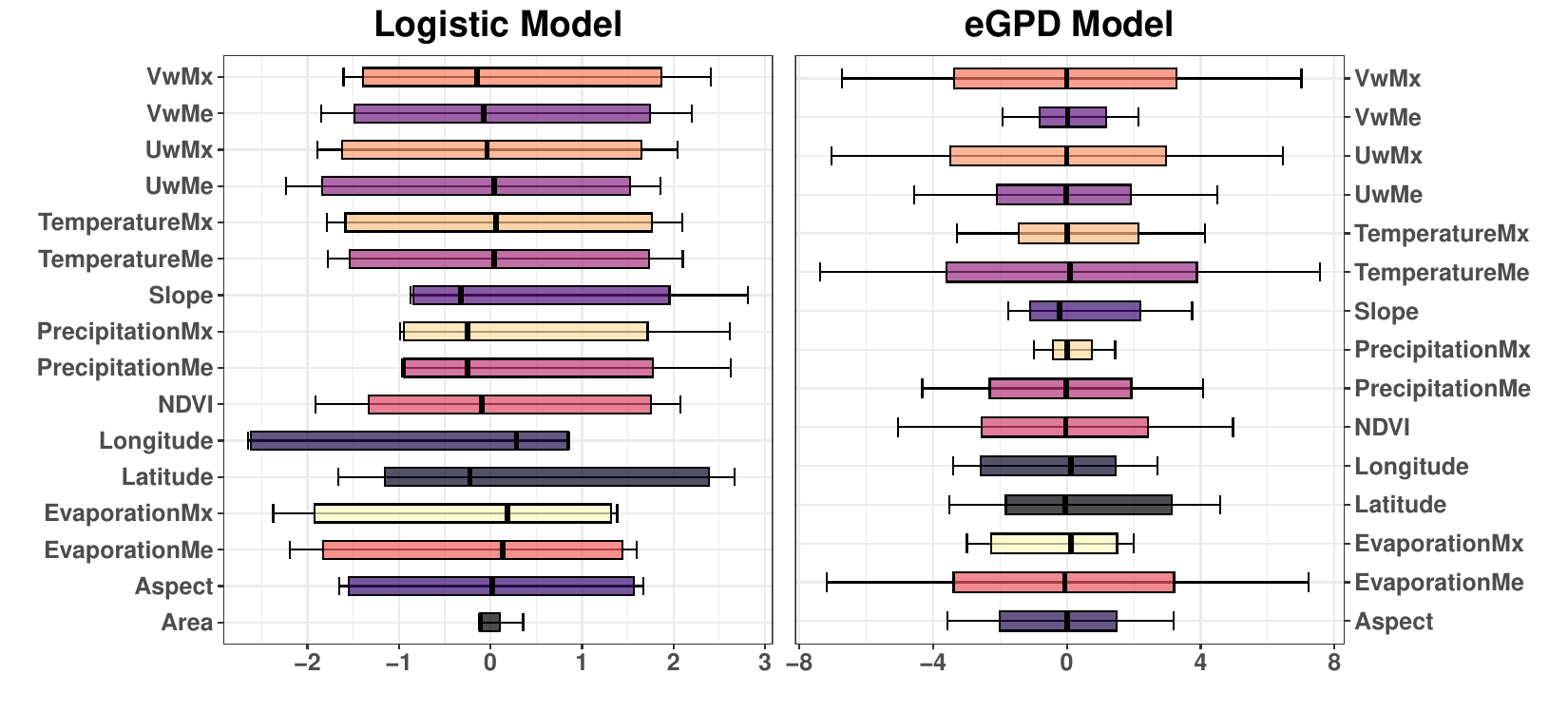}
    \caption{Boxplots of estimated contribution scores $\mbox{CS}_{i,(\bm s, t)}$ for occurrence probability $p_0(\bm s, t)$ [unitless] (left) and  conditional spread severity $\sigma(\bm s,t)$ [km] (right). Scores are estimated and pooled over all space-time locations $(\bm s, t)$, and displayed for each covariate $i=1,\dots,d$ (rows).}
    \label{fig:Boxplot}
\end{figure}


In Figure \ref{fig:Boxplot}, we present boxplots of the estimated contribution scores for both $p_0(\bm s, t)$ and $\sigma(\bm s, t)$, pooled over all space-time locations. For the occurrence probability $p_0(\bm s,t)$, estimates from the model suggest that all covariates, except the area $a( \bm s)$, have approximately equivalent importance. For the eGPD model, the five most important covariates for estimating the conditional spread severity $\sigma(\bm s, t)$ are the monthly mean of temperature (TemperatureMe) and evaporation (EvaporationMe), monthly maxima of both wind speed components (UwMx and VwMx), and NDVI. We note that characteristics of the land surface, including slope and aspect, have a relatively less important impact on the conditional distribution of the monthly radial spread, given a wildfire occurrence. Such difference between the covariate effects retrieved for the occurrence probability and the eGPD models could be interpreted with both terrain and climatic influences being required to initiate the fire, whereas the spreading process may be mostly controlled by meteorological conditions and vegetation types. Interestingly, we find that monthly maxima of precipitation, temperature, and evaporation play a much weaker role than their corresponding monthly means, suggesting that it is typical climate conditions that drive wildfire spread, rather than climatic extremes.

\section{Conclusion}
\label{sec:Conclusions}

We have developed a hybrid statistical deep-learning regression model for burnt areas from Australian wildfires. Our approach uses a semi-parametric regression model with graph convolutional neural networks and the extended generalized Pareto distribution, and enables us to accurately model extreme wildfire spread observed over an irregular spatial domain. Using a novel dataset of Australian wildfires spanning June 1999 to December 2019, we analyze monthly burnt area over irregularly arranged polygons of approximately equal population density. We assess the proportional hazard attributed to wildfires in each polygon and identify those regions most susceptible to wildfires. We also quantify the relative importance of the considered covariates in the estimation of the distribution of wildfire occurrence and spread.

Our deep learning framework facilitates fantastic model fits by capturing the spatial structure exhibited by the covariates. We adopt a relatively simple, but powerful, convolution-based graph neural network, and so future work may include exploiting alternative graph neural networks to further improve model performance; for a recent review of such methods, see, for example, \cite{zhou2020graph}. Whilst we focus on capturing only spatial structure in our data, graph neural networks that capture space-time structures do exist and their efficacy could be explored; one choice could be the recurrent graph convolutional neural network of \cite{seo2018structured}, which extends the regular convolutional long short-term memory network \citep{shi2015convolutional}, to a graphical setting. 

A further methodological pursuit is to allow the parameters in the adjacency matrix, $A$ in \eqref{eq:adj}, to be jointly estimable with the neural network weights, rather than optimising $A$ using a grid search, or to consider different forms for $A$. We use the great-circle distance metric in \eqref{eq:adj} and, whilst we found this works well for our data, it is not designed to measure distances between polygons (which we treat as point locations). More appropriate forms for $A$ could be chosen which account for the irregular lattice structure of our data, e.g., with elements $A[ij]$ having more weight if regions $\bm s_i$ and $\bm s_j$ share a longer border. Alternatively, we could exploit graph neural networks that benefit from learnable edge weights
 \cite[see, e.g.,][]{jiang2020co}, where the initial specification for $A$ may be somewhat arbitrary. 

Finally, we note that, whilst our framework exploits spatial structure in the covariates it does not explicitly account for dependence in the burnt area response $Y(\bm s, t)$. Hence, it cannot be used to quantify the joint risk of wildfires at different space-time locations $(\bm s, t)$ or provide estimates of high quantiles for aggregates of wildfire burnt area over space-time regions. This limitation can be solved by couching our deep eGPD regression model within a specialised spatial extremal dependence model \citep[see, e.g.,][]{Davison.Huser:2015,huser2020advances}, which we leave as a further consideration.
\section{Acknowledgements}
Cisneros, Richards, and Huser were supported by funding from the King Abdullah University of Science and Technology (KAUST) Office of Sponsored Research (OSR) under Awards No. OSR-CRG2020-4394 and ORA-2022-5336. Dahal and Lombdardo were supported by funding from the King Abdullah University of Science and Technology (KAUST) Competitive Research Grant (CRG) under award No. URF/1/4338-01-01. Support from the KAUST Supercomputing Laboratory is gratefully acknowledged.
\baselineskip 20pt
\bibliographystyle{apalike}
\bibliography{references}

\baselineskip 10pt
\newpage


\end{document}